# Size Effect of Negative Capacitance State and Subthreshold Swing in Van der Waals Ferrielectric Field-Effect Transistors

Anna N. Morozovska[1*], Eugene A. Eliseev[2], Yulian M. Vysochanskii[3], Sergei V. Kalinin[4†], and Maksym V. Strikha[5,6‡]

[1] Institute of Physics, National Academy of Sciences of Ukraine,
Pr. Nauky 46, 03028 Kyiv, Ukraine,

[2] Frantsevich Institute for Problems in Materials Science, National Academy of Sciences of Ukraine, 3, str. Omeliana Pritsaka, 03142 Kyiv, Ukraine

[3] Institute of Solid State Physics and Chemistry, Uzhhorod University,
88000 Uzhhorod, Ukraine

[4] Department of Materials Science and Engineering, University of Tennessee,
Knoxville, TN, 37996, USA

[5] Taras Shevchenko National University of Kyiv, Faculty of Radiophysics, Electronics and Computer Systems, Pr. Akademika Hlushkova 4g, 03022 Kyiv, Ukraine,

[6] V. Lashkariov Institute of Semiconductor Physics, National Academy of Sciences of Ukraine, Pr. Nauky 41, 03028 Kyiv, Ukraine

**Abstract**

Analytical calculations corroborated by the finite element modelling show that thin films of Van der Waals ferrielectrics covered by a 2D-semiconductor are promising candidates for the controllable reduction of the dielectric layer capacitance due to the negative capacitance (NC) effect emerging in the ferrielectric film. The NC state is conditioned by energy-degenerated poly-domain states of the ferrielectric polarization induced in the films under incomplete screening conditions in the presence of a dielectric layer. Calculations performed for the FET-type heterostructure "ferrielectric $CuInP_2S_6$ film – 2D-$MoS_2$ single-layer – $SiO_2$ dielectric layer" reveal the pronounced size effect of the multilayer capacitance. Derived analytical expressions for the electric polarization and multilayer capacitance allow to predict the thickness range of the dielectric layer and ferrielectric film for which the NC effect is the most pronounced in various Van der Waals ferrielectrics, and the corresponding subthreshold swing becomes much less than the Boltzmann's limit. Obtained results can be useful for the size and temperature control of the NC effect in the steep-slope ferrielectric FETs.

---

[*] corresponding author, e-mail: anna.n.morozovska@gmail.com
[†] Corresponding author: sergei2@utk.edu
[‡] Corresponding author: maksym.strikha@gmail.com

## I. INTRODUCTION

The stabilization of ferroelectric thin films in the state of negative differential capacitance (NC) was revealed experimentally and it was demonstrated [1, 2] that the total capacitance of a double-layer capacitor, made of a dielectric layer and a ferroelectric film, can be greater than it would be for the single-layer capacitor filled with the dielectric of same thickness as used in the double-layer capacitor. The ferroelectric in the NC state placed in the gate stack of the field-effect transistor (FET) is regarded very promising for low power consumption electronic devices, chips and logic circuits [3]. NC-FETs with the channel made from a 2D-semiconductor have all potential opportunities for ultra-low power consumption devices applications, at that they can combine ultra-steep switching ability of NC-FETs introduced by ferroelectric materials and superior gate electrostatic controllability provided by 2D-semiconductors with high carrier mobility [4].

Experimental observations of the NC effect in ferroelectric multilayer capacitors [5, 6, 7, 8] are often, but there are a few works devoted to the semi-analytical description of the NC state appearance conditions, which also consider the influence of the domain structure in the ferroelectric layer on the NC state (see e.g., Refs. [9, 10, 11, 12]). The analytical description is rare because it is very difficult to find the analytical conditions of the NC state appearance and stability under the possible presence of the domains in the ferroelectric film allowing for their morphology and size effects.

Appeared that thin films of Van der Waals ferrielectrics, such as $CuInP_2S_6$ (CIPS), are well-suitable for the analytical control of the NC effect [13], which has been observed experimentally in the layered materials [14, 15, 16]. The reason why the analytical control of the NC effect is easily possible in CIPS films, is the 8-th order thermodynamic potential of CIPS, which can contain four wells flattening in the vicinity of the "critical end point" (CEP) and "bicritical end point" (BEP) [17, 18, 19]. Recent analytical calculations [13] predict that a high-performance NC state appears in CIPS film when the flat wells split into a very wide plateau in the presence of the dielectric layer and for the definite range of tensile strains originated from the CIPS film-substrate lattice mismatch. However, results [13] are obtained in the single-domain approximation for the CIPS film.

The subthreshold swing $S$ is a fundamental characteristic of a field-effect transistor (FET). The parameter $S$ shows how many times it is necessary to increase the gate voltage in the subthreshold region to achieve an increase in the drain current by an order of magnitude. For a high-quality transistor with a large value of the positive gate capacitance, the limiting value of $S$ is equal to $\frac{k_B T}{e}\ln 10$, where $e$ is the electron charge, $k_B$ is the Boltzmann constant and $T$ is the temperature in Kelvins. At room temperature $S \approx 60$ mV/decade. The importance of the subthreshold swing reduction follows from the fact that its smallest limiting value determine the minimal possible operating voltage of the transistor supply. Therefore reducing $S$ below the fundamental limit can open great prospects for further reduction of the power consumption, creation of miniature steep-slope FETs and related logic devices [20]. Indeed, Wang et al. [14] demonstrated the steep-slope NC-FETs with the channel made of two-dimensional (2D) MoS$_2$ and the CIPS thin film of thickness less than 20 nm as a gate oxide. The CIPS-2D-MoS$_2$ NC-FET has a subthreshold swing much less than the Boltzmann's limit. Dey et al. [15] revealed that the NC state of CIPS film can stabilize logic states in tunnel FETs. Chi et al. [16] demonstrated that the elastic strain of the CIPS film affects strongly the working performances of the Van der Waals NC-FETs.

In this work we perform analytical calculations in the framework of Landau-Ginzburg-Devonshire (LGD) approach and finite element modelling (FEM) of the electric and elastic fields, polarization and capacitance in the FET-type heterostructure "CIPS film – 2D-MoS$_2$ single-layer – SiO$_2$ dielectric layer". We paid special attention to the domain stripes appearance in the CIPS film and reveal that the NC state is conditioned by energy-degenerated poly-domain states of the ferrielectric polarization. Using a direct variational method, we derived analytical expressions for the ferrielectric polarization average value, amplitude and period of the emergent domain structure, which allows to calculate the polarization and multi-layer capacitance phase diagrams. The analytical expressions allow to predict the thickness range of the dielectric layer and ferrielectric film corresponding to the most pronounced NC effect and to the subthreshold swing much less than the Boltzmann's limit. The paper is structured as follows: **Section II** contains the formulation of the problem, basic equations, assumptions and analytical solutions. **Section III** contains the discussion and analysis of analytical and FEM results. **Section IV** is a summary. Calculation details are listed in **Supplemental Materials** [21].

## II. THEORETICAL DESCRIPTION
### A. The Problem Formulation

Let us consider a multilayer capacitor consisting of the top electrode, the CIPS film of thickness $h$, the single-layer MoS$_2$, which is treated as an infinitely thin 2D-semiconductor with the effective screening length $\lambda$, the dielectric SiO$_2$ layer of thickness $d$, and the bottom electrode. The multilayer capacitor is shown in **Fig. 1(a)**. The spontaneous ferrielectric polarization of the CIPS layer is regarded directed along the polar axis $z$, and so $\vec{P} = (0,0,P_S)$. The dependence of the spontaneous polarization $P_S$ on temperature $T$ and mismatch strain $u_m$ for a CIPS film covered by ideally conducting electrodes is shown in **Fig. 1(b)**. Here PE is the paraelectric phase, where $P_S = 0$, FI1 and FI2 are the ferrielectric states with higher and lower $P_S$. It is seen from **Fig. 1(b)** that tensile strains favor the appearance of the PE phase and FI2 state, while compressive strains favor the stability of the FI1 state.

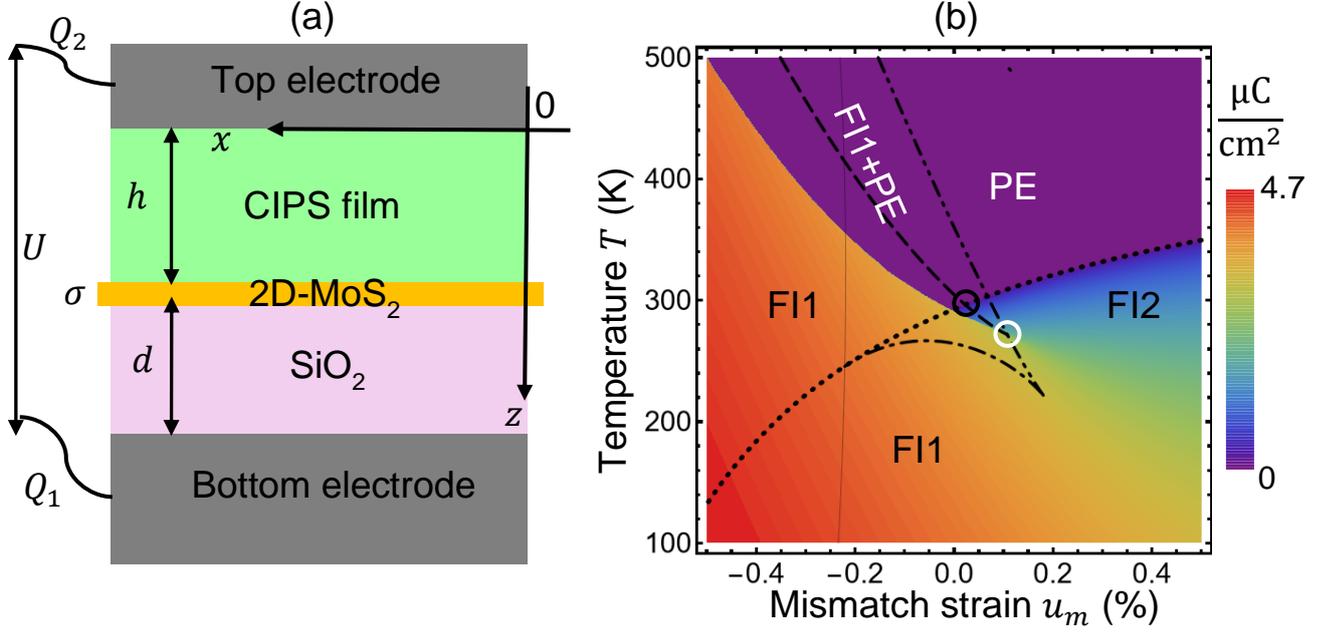

**Figure 1.** **(a)** The multilayer capacitor consisting of the top electrode, the CIPS film of thickness $h$, the 2D-MoS$_2$ single-layer, the dielectric SiO$_2$ layer of thickness $d$, and the bottom electrode. **(b)** The dependence of the spontaneous polarization $P_S$ on temperature $T$ and mismatch strain $u_m$ for a CIPS film covered by ideally conducting electrodes. PE is the paraelectric phase, FI1 and FI2 are the ferrielectric states with higher and lower $P_S$ values. CEP and BEP are the critical and bicritical end points, marked by the black and white circles, respectively. Color scale is the absolute value of $P_S$ in the deepest potential well of the LGD free energy. The part **(b)** is reprinted from Ref. [22].

The connection between the electric displacement $\vec{D}$ and the electric field, $\vec{E}$, in the ferrielectric ($f$) CIPS and the dielectric($d$) SiO$_2$ layers are:

$$\vec{D}^f = \varepsilon_0 \varepsilon_b \vec{E}^f + \vec{P}, \quad \vec{D}^d = \varepsilon_0 \varepsilon_d \vec{E}^l, \qquad (1)$$

where $\varepsilon_0$ is a universal dielectric constant, $\varepsilon_b$ is a relative background permittivity of the CIPS layer, $\varepsilon_d$ is the relative dielectric permittivity of the dielectric SiO$_2$ layer. To find the electric field $\vec{E} = -\nabla \varphi$, one should solve the following equations for the electrostatic potential in the ferrielectric and dielectric layers:

$$\varepsilon_0 \varepsilon_b \left( \frac{\partial^2}{\partial x^2} + \frac{\partial^2}{\partial y^2} + \frac{\partial^2}{\partial z^2} \right) \varphi_f = \left( \frac{\partial P_1}{\partial x} + \frac{\partial P_2}{\partial y} + \frac{\partial P_3}{\partial z} \right), \quad 0 < z < h, \qquad (2a)$$

$$\varepsilon_0\varepsilon_d\left(\frac{\partial^2}{\partial x^2}+\frac{\partial^2}{\partial y^2}+\frac{\partial^2}{\partial z^2}\right)\varphi_d=0, \qquad h<z<h+d. \qquad (2b)$$

The electrostatic potential is continuous at the interfaces $z=0$, $z=h$ and $z=h+d$, and the normal component of the electric displacement differs by the value of the surface charge density $\sigma$ at the CIPS-2D-MoS$_2$-SiO$_2$ interface $z=h$. Taking this into account, the boundary conditions for Eqs.(2) are:

$$\varphi_d(x,y,h+d)=U, \quad \varphi_f(x,y,h)=\varphi_d(x,y,h), \quad \varphi_f(x,y,0)=0, \qquad (3a)$$

$$D_z^d(x,y,h)-D_z^f(x,y,h)=\sigma. \qquad (3b)$$

Hereinafter $\sigma$ is the charge density of the 2D-MoS$_2$ layer.

The Landau-Ginzburg-Devonshire-Khalatnikov equation, which determines the ferrielectric polarization $P_3$ of the CIPS layer has the form [23]:

$$\Gamma\frac{d}{dt}P_3+\tilde{\alpha}(T)P_3+\tilde{\beta}P_3^3+\tilde{\gamma}P_3^5+\tilde{\delta}P_3^7-g_{33}\frac{\partial^2}{\partial z^2}P_3-g_{44}\left(\frac{\partial^2}{\partial x^2}+\frac{\partial^2}{\partial y^2}\right)P_3=E_3^f. \qquad (4)$$

Here $\Gamma$ is the Khalatnikov coefficient, $\tilde{\alpha}$, $\tilde{\beta}$, $\gamma$, and $\delta$ are Landau expansion coefficients, $g_{ij}$ are polarization gradient coefficients, and $E_3^f$ is the electric field, which satisfies Eq.(2) with the boundary conditions (3). The polarization satisfies the third kind boundary conditions at the film surfaces ($z=0$ and $z=h$), which we regard "natural" for the sake of maximal stability of the polar state, $\left(\frac{\partial P_3}{\partial z}\right)\Big|_{z=0,h}=0$. The Landau expansion coefficients in Eq.(4) are renormalized by electrostriction coupling as:

$$\tilde{\alpha}(T)=\alpha_T(T-T_C)-2\frac{Q_{13}(s_{22}-s_{12})+Q_{23}(s_{11}-s_{12})}{s_{11}s_{22}-s_{12}^2}u_m, \qquad (5a)$$

$$\tilde{\beta}=\beta+2\frac{Q_{23}^2s_{11}-2Q_{13}Q_{23}s_{12}+Q_{13}^2s_{22}}{s_{11}s_{22}-s_{12}^2}-4\frac{s_{22}Z_{133}+s_{11}Z_{233}-s_{12}(Z_{133}+Z_{233})}{s_{11}s_{22}-s_{12}^2}u_m, \qquad (5b)$$

$$\tilde{\gamma}=\gamma+6\frac{Q_{13}s_{22}Z_{133}+Q_{23}s_{11}Z_{233}-Q_{13}s_{12}Z_{233}-Q_{23}s_{12}Z_{133}}{s_{11}s_{22}-s_{12}^2}, \qquad (5c)$$

$$\tilde{\delta}=\delta+4\frac{s_{22}Z_{133}^2-2s_{12}Z_{133}Z_{233}+s_{11}Z_{233}^2}{(s_{11}s_{22}-s_{12}^2)}. \qquad (5d)$$

In Eq.(5a) $T$ is the temperature and $T_C$ is the Curie temperature of a bulk CIPS. The terms in Eqs.(1), which are proportional to the mismatch strain $u_m$ and/or combination of the second order electrostriction coefficients $Q_{ij}$, higher order electrostriction coefficients $Z_{ijk}$ and elastic compliances $s_{ij}$, originate from the elastic strains coupling with the electrostriction [24]. The values of $T_C$, $\alpha_T$, β, $\gamma$ and δ, $g_{ij}$, $Q_{ij}$ and $s_{ij}$, and $\lambda$ are listed in **Table AI** in **Appendix A** [21].

## B. Analytical Solutions for the Ferrielectric Polarization, Potential and Capacitance

The solutions for the polarization and electric field could be essentially simplified if we assume the linear dependence of the surface charge $\sigma$ on the electric potential $\varphi(x, y, z)$ at $z = h$ [25, 26]:

$$\sigma(x, y) = -\varepsilon_0 \frac{\varphi(x, y, h)}{\lambda}. \tag{6}$$

Here $\lambda$ is the effective screening length of the single layer MoS$_2$, which is treated as the 2D-semiconductor.

The CIPS film can be in the poly-domain or single-domain ferrielectric states with low- or high-polarization, or in the depolarized paraelectric phase. The approximate analytical expression for polarization, listed below, is valid in the paraelectric (**PE**) phase, in the "shallow" polydomain ferrielectric states (**PDFI1** and **PDFI2**) near the PE phase boundary, where the ferrielectric polarization is harmonically modulated, and in the "deep" single-domain ferrielectric states (**SDFI1** and **SDFI2**). Due to the 8-th order LGD free energy, the rippled state (**PDFI+SDFI**), where the domain modulation coexists with the homogeneously polarized state, can be stable for a wide range of parameters. The case of the "deep" PDFI states with an-harmonically modulated polarization should be considered numerically. The approximate expression for $P_3$ in the PE phase, SDFE and shallow PDFI states has the following form:

$$P_3 \approx \bar{P} + 2A \cos(kx) \left[\cos(\xi z) + \frac{\xi \sin(\xi h)}{\mu \sinh(\mu h)} \cosh(\mu z)\right]. \tag{7a}$$

Here $\bar{P}$ is the polarization $P_3$ averaged over the thickness of the CIPS film. $A$ is the amplitude of the domain modulation. $\bar{P} = 0$ in the PDFI states and in the PE phase for $U = 0$. $A = 0$ in the PE phase and in the pure SDFI state. Both $\bar{P}$ and A are nonzero in the coexisting region of PDFI and SDFI states.

In **Appendix C** [21] we derive the approximate expressions for the positive size-dependent function $\xi$ and positive size-independent constant $\mu$:

$$\xi \approx \frac{k}{\sqrt{\left(\frac{2}{\pi}kh\right)^2 + \left(\frac{\varepsilon_d k}{\tanh(kd)} + \frac{1}{\lambda}\right)\frac{h}{\varepsilon_b}}}, \qquad \mu \approx \sqrt{\frac{1}{\varepsilon_0 \varepsilon_b g_{33}}}. \tag{7b}$$

The parameter $\sqrt{\varepsilon_0 \varepsilon_b g_{33}} \sim 0.3$ nm is smaller than the lattice constant and has a sense of a correlation length in the polar direction z. At the PE-PDFI boundary the wavenumber $k$ can be found from the cumbersome equation (C.8c), which form is determined by the boundary conditions (see **Appendix C** [21] for details). The extremal, i.e., the "threshold", value of $k_{cr}$ can be written in a relatively simple form:

$$k_{cr} \approx \frac{\pi}{2h}\sqrt{\frac{h}{h_{cr}} - 1}. \tag{7c}$$

The PDFI state can exist in the temperature range $T < T_{PDFI}$, and the temperature of the state instability is

$$T_{PDFI} = \tilde{T}_C - \frac{g_{44}}{\alpha_T \varepsilon_b} \frac{\pi^2}{4h} \left( 2\frac{\varepsilon_b}{l_{cx}} - \frac{\varepsilon_b}{h} - \frac{\varepsilon_d}{d} - \frac{1}{\lambda} \right), \quad (7d)$$

where $\tilde{T}_C = T_C + \frac{2}{\alpha_T} \frac{Q_{13}(s_{22}-s_{12})+Q_{23}(s_{11}-s_{12})}{s_{11}s_{22}-s_{12}^2} u_m$. The small parameter $l_{cx} = \frac{\pi}{2}\sqrt{\varepsilon_0 \varepsilon_b g_{44}} \sim 0.5$ nm has the sense of a correlation length in the transverse direction x.

From Eqs. (7c), the PE-PDFI transition can occur only if the CIPS thickness $h$ is bigger than the critical thickness, $h > h_{cr}$. For $h < h_{cr}$ the PE-SDFI transition may happen at the temperature $T_{SDFI} = \tilde{T}_C - \frac{1}{\alpha_T \varepsilon_0 \left(\varepsilon_b + \frac{h}{d}\varepsilon_d + \frac{h}{\lambda}\right)}$. For $h \gg h_{cr}$ the transition to the SDFI state gradually happens with increase in $h$. The critical thickness of ferroelectric layer depends on the dielectric layer thickness and screening length, as

$$h_{cr} = \varepsilon_b \left( \frac{\varepsilon_b}{l_{cx}} - \frac{\varepsilon_d}{d} - \frac{1}{\lambda} \right)^{-1}. \quad (7e)$$

The condition (7e) means that the critical thickness of a ferrielectric film exists, which value is determined by the thickness of the dielectric layer, screening and correlation lengths. From Eq.(7e) the wave vector $k_{cr}$ is real only under the following condition

$$\frac{\varepsilon_b}{h} + \frac{\varepsilon_d}{d} + \frac{1}{\lambda} < \frac{\varepsilon_b}{l_{cx}}. \quad (7f)$$

For $h \geq h_{cr}$ and $T < T_{PDFI}$ the minimal domain period $D_{cr}$ is given by expression $D_{cr} = \frac{2\pi}{k_{cr}}$. The dependence of $D_{cr}$ on the film thickness $h$ calculated for several values of $\lambda$ and $d$ is shown in **Fig. 2**. The value of $D_{cr}$ linearly increases from 1.5 nm to 5.5 nm with increase in $h$ from 2 nm to 20 nm; also, it slightly increases with decrease in $d$ (see different curves in **Fig. 2(a)**) or $\lambda$ (see different curves in **Fig. 2(b))**.

Note that the increase in $D_{cr}$ can lead to the appearance and increase of the unipolarity degree of the CIPS films. Namely, the unipolarity degree increases once the film x-size $L$ becomes smaller than $D_{cr}$, and the film should be treated as a thin plate. In the case $L < \frac{D_{cr}}{2}$ the single-domain state of the plate can be caused by the spatial confinement in the transverse {x,y} directions. However, the full screening conditions in the x and y directions are required for support of the CIPS plate unipolarity. Since small $L$ values are required for usage of CIPS plates in the FETs with short channels, the SDFI state induced by the transverse confinement is of special importance. For the case of ultra-short channel ($L \leq 10$ nm) the gate-all-around architecture of FET can be used to provide the high degree of screening of the CIPS plate in the transverse directions.

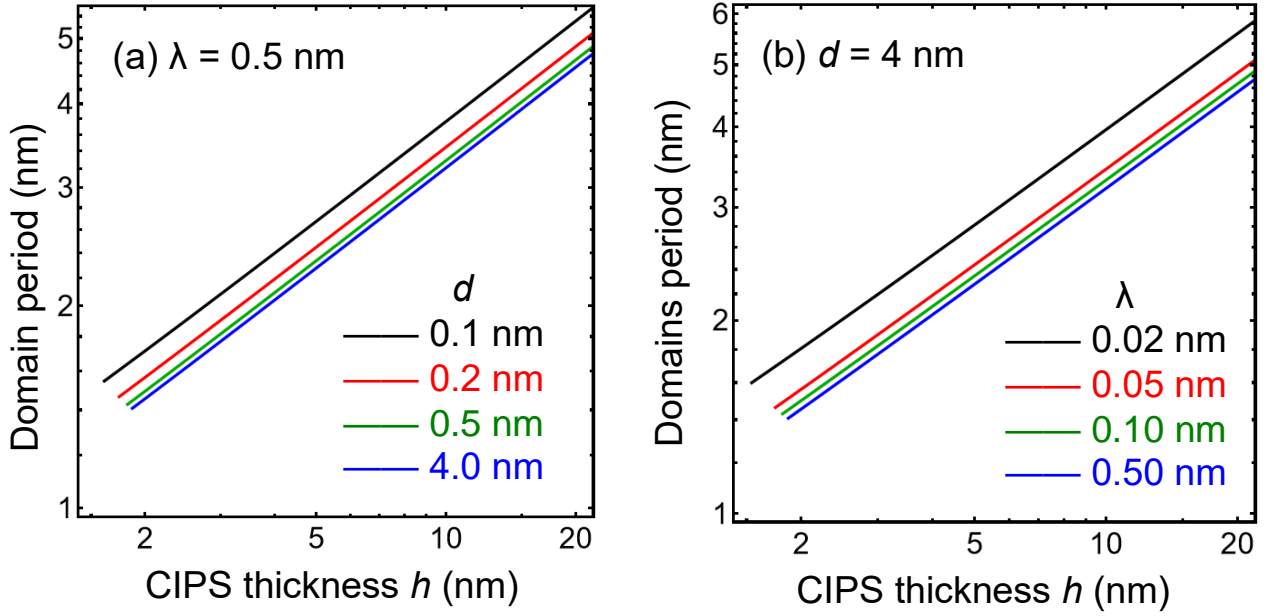

**Figure 2.** The dependence of the domain period $D_{cr}$ on the CIPS film thickness $h$ calculated for the screening length $\lambda = 0.5$ nm and different thickness of the dielectric layer $d = 0.1, 0.2, 0.5$ and $4$ nm (black, red, green and blue curves, respectively) **(a)**; for $d = 4$ nm and $\lambda = 0.02, 0.05, 0.1$ and $0.5$ nm (black, red, green and blue curves) **(b)**. Plots are calculated for $T = 293$ K, misfit strain $u_m = 0.3\%$, $\varepsilon_d = 3.9$.

Coupled algebraic equations, which determine $\bar{P}$ and $A$ values in Eq.(7a), have the form:

$$(\alpha_P + 3\tilde{\beta}A^2)\bar{P} + (\tilde{\beta} + 10\tilde{\gamma}A^2)\bar{P}^3 + (\tilde{\gamma} + 42\tilde{\delta}A^2)\bar{P}^5 + \tilde{\delta}\bar{P}^7 = -\frac{\varepsilon_d U}{d\,h\left(\frac{1}{\lambda}+\frac{\varepsilon_b}{h}+\frac{\varepsilon_d}{d}\right)}, \quad (8a)$$

$$(\alpha_A + 3\tilde{\beta}\,\bar{P}^2)A + \frac{9}{4}\tilde{\beta}\,A^3 = 0. \quad (8b)$$

The renormalized size-dependent coefficients $\alpha_P$ and $\alpha_A$ are given by expressions:

$$\alpha_P = \tilde{\alpha} + \frac{1}{\varepsilon_0 h\left(\frac{1}{\lambda}+\frac{\varepsilon_b}{h}+\frac{\varepsilon_d}{d}\right)}, \quad \alpha_A = \tilde{\alpha} + g_{44}k^2 + \frac{1}{\varepsilon_0\varepsilon_b}\frac{1+\varepsilon_0\varepsilon_b g_{33}k^2}{1+\left(\frac{2}{\pi}kh\right)^2+\left(\frac{\varepsilon_d k}{\tanh(kd)}+\frac{1}{\lambda}\right)\frac{h}{\varepsilon_b}}. \quad (8c)$$

From Eqs.(8b), the amplitude $A$ is zero in the PE phase and in the SDFI state, and $A \approx \frac{2}{3}\sqrt{\frac{-\alpha_A - 3\tilde{\beta}\,\bar{P}^2}{\tilde{\beta}}}$ in the shallow PDFI state and in the rippled PDFI+SDFI state.

In the case of small electric potential $U$ the solution of Eq.(8a) can be linearized over $U$ and acquires the form:

$$\bar{P} = \bar{P}_0 + \delta P, \quad \delta P = -\frac{\varepsilon_d U}{d\,h\left(\frac{1}{\lambda}+\frac{\varepsilon_b}{h}+\frac{\varepsilon_d}{d}\right)}\left(\tilde{\alpha}_R + 3\tilde{\beta}_R\bar{P}_0^2 + 5\tilde{\gamma}_R\bar{P}_0^4 + 7\tilde{\delta}\bar{P}_0^6\right)^{-1}. \quad (9a)$$

The expressions for renormalized coefficients $\tilde{\alpha}_R$, $\tilde{\beta}_R$ and $\tilde{\gamma}_R$ depends on the film phase state, namely:

$$\tilde{\alpha}_R = \begin{cases} \alpha_P, & \text{for the PE and SDFI states,} \\ -\frac{4}{3}\alpha_A + \alpha_P, & \text{for the PDFI state.} \end{cases} \quad (9b)$$

$$\tilde{\beta}_R = \begin{cases} \tilde{\beta}, & \text{for the PE and SDFI states,} \\ \tilde{\beta} - \frac{40\tilde{\gamma}}{9\tilde{\beta}}\alpha_A, & \text{for the PDFI state.} \end{cases} \quad (9c)$$

$$\tilde{\gamma}_R = \begin{cases} \tilde{\gamma}, & \text{for the PE and SDFI states,} \\ \tilde{\gamma} - \frac{28\tilde{\delta}}{3\tilde{\beta}}\alpha_A, & \text{for the PDFI state.} \end{cases} \qquad (9d)$$

The spontaneous polarization $\bar{P}_0 = 0$ in the PE phase (where also $A = 0$) and in the shallow PDFI phase (where $A^2 \approx -\frac{4}{9\tilde{\beta}}\alpha_A$). $\bar{P}_0$ is nonzero and satisfies the nonlinear equation in the pure SDFI and rippled PDFI+SDFI states:

$$\alpha_P + \tilde{\beta}\bar{P}_0^2 + \tilde{\gamma}\bar{P}_0^4 + \tilde{\delta}\bar{P}_0^6 = 0. \qquad (9e)$$

In **Appendices A** and **B** [21] we derived the analytical expressions for the interface potential $\varphi(h)$ and the surface charge $\sigma$, charges at upper and lower electrodes, $Q_1$ and $Q_2$, and the "discharge" $Q$:

$$\varphi(h) = \frac{1}{\frac{1}{\lambda} + \frac{\varepsilon_b}{h} + \frac{\varepsilon_d}{d}} \frac{\bar{P}}{\varepsilon_0} + \frac{\varepsilon_d U}{d\left(\frac{1}{\lambda} + \frac{\varepsilon_b}{h} + \frac{\varepsilon_d}{d}\right)}, \qquad (10a)$$

$$\sigma = -\frac{\bar{P}}{\frac{1}{\lambda} + \frac{\varepsilon_b}{h} + \frac{\varepsilon_d}{d}} - \frac{\varepsilon_0 \varepsilon_d U}{d\lambda\left(\frac{1}{\lambda} + \frac{\varepsilon_b}{h} + \frac{\varepsilon_d}{d}\right)}, \qquad (10b)$$

$$Q_1 = \frac{\varepsilon_0 \varepsilon_d \varepsilon_b U}{d\, h\left(\frac{1}{\lambda} + \frac{\varepsilon_b}{h} + \frac{\varepsilon_d}{d}\right)}\left(1 + \frac{h}{\varepsilon_b \lambda}\right) - \frac{\varepsilon_d}{d\left(\frac{1}{\lambda} + \frac{\varepsilon_b}{h} + \frac{\varepsilon_d}{d}\right)}\bar{P}, \quad Q_2 = \frac{-\varepsilon_0 \varepsilon_d \varepsilon_b U}{d\, h\left(\frac{1}{\lambda} + \frac{\varepsilon_b}{h} + \frac{\varepsilon_d}{d}\right)} + \frac{\left(\frac{d}{\lambda} + \varepsilon_d\right)}{d\left(\frac{1}{\lambda} + \frac{\varepsilon_b}{h} + \frac{\varepsilon_d}{d}\right)}\bar{P}, \qquad (10c)$$

$$Q = \frac{Q_1 - Q_2}{2} = -\frac{\left(\varepsilon_d + \frac{d}{2\lambda}\right)}{d\left(\frac{1}{\lambda} + \frac{\varepsilon_b}{h} + \frac{\varepsilon_d}{d}\right)}\bar{P} + \frac{\varepsilon_0 \varepsilon_d \varepsilon_b U}{d\, h\left(\frac{1}{\lambda} + \frac{\varepsilon_b}{h} + \frac{\varepsilon_d}{d}\right)}\left(1 + \frac{h}{2\lambda\, \varepsilon_b}\right). \qquad (10d)$$

The identity $Q_1 + Q_2 + \sigma = 0$, which follows from expressions (10), is consistent with the electroneutrality condition.

From Eqs.(9) and (10), the evident form of the effective differential capacitance of the multilayer is

$$C_{eff} = \frac{dQ}{dU} \approx \frac{\varepsilon_0 \varepsilon_d}{d} \frac{\frac{\varepsilon_a}{h} + \frac{1}{2\lambda}}{\frac{\varepsilon_d}{d} + \frac{1}{\lambda} + \frac{\varepsilon_a}{h}}. \qquad (11a)$$

Here the apparent dielectric permittivity of the CIPS layer $\varepsilon_a$ is introduced as:

$$\varepsilon_a = \frac{1}{\varepsilon_0 \alpha_0} + \varepsilon_b, \qquad \alpha_0 = \tilde{\alpha}_R + 3\tilde{\beta}_R \bar{P}_0^2 + 5\tilde{\gamma}_R \bar{P}_0^4 + 7\tilde{\delta}\bar{P}_0^6. \qquad (11b)$$

The denominator in Eq.(11a) is the sum of "effective" differential capacitances of the dielectric layer $\frac{\varepsilon_d}{d}$, 2D-semiconducor, $\frac{1}{\lambda}$, and the ferrielectric film, $\frac{\varepsilon_a}{h}$. Note, that expressions (11b) are basic for understanding of the NC state appearance in ferroelectrics, since $\varepsilon_b$ is always positive, but $\alpha_0$ can be negative for small polarizations $\bar{P}$ (see e.g. Ref.[1]), which can lead to the negative values of $\varepsilon_a$.

Since the capacitance of the reference capacitor made of the SiO$_2$ layer with thickness $d$ is $C_d = \frac{\varepsilon_0 \varepsilon_d}{d}$, the expression for the relative effective capacitance acquires the form:

$$\Delta C = \frac{C_d - C_{eff}}{C_d} = \frac{\frac{1}{2\lambda} + \frac{\varepsilon_d}{d}}{\frac{1}{\lambda} + \frac{\varepsilon_a}{h} + \frac{\varepsilon_d}{d}}. \qquad (12)$$

The effect of the NC corresponds to the case $\Delta C < 0$, and can be realized if $\varepsilon_a < 0$. We can estimate the apparent capacitance of the "CIPS + MoS$_2$" layer $C_f$ as the solution of equation $\frac{1}{C_f} = \frac{1}{C_{eff}} - \frac{1}{C_d}$, and introduce its relative value as

$$\Delta C_f = \frac{C_f}{C_d} = \frac{C_{eff}}{C_d - C_{eff}} = \frac{C_d}{C_d - C_{eff}} - 1 = \frac{1}{\Delta C} - 1 = \frac{\frac{1}{2\lambda} + \frac{\varepsilon_a}{h}}{\frac{1}{2\lambda} + \frac{\varepsilon_d}{d}}. \qquad (13)$$

## IV. RESULTS AND DISCUSSION
### A. Results of Analytical Calculations

The dependences of the average spontaneous polarization $\bar{P}$, relative effective capacitance $\Delta C$, and the CIPS layer apparent capacitance $\Delta C_f$ on the relative sizes $h/\lambda$ and $d/\lambda$ are shown in **Fig. 3(a), 3(b)** and **3(c)**, respectively. To reach the NC state in the maximal ranges of sizes and temperatures, we apply a relatively small tensile misfit strain, $u_m = 0.3\%$, to the CIPS film. The tensile strains favor the appearance of the PE phase (with $\bar{P} = 0$) and/or the FI2 state (with small $\bar{P}$) (see **Fig. 1(b)**), for which the region of NC effect appears maximal in comparison with zero and compressive strains.

The black solid curve in **Fig. 3(a)** is the boundary separating the PE phase and the PDFI state; the black dotted curve is the boundary of the SDFI state absolute instability; and the black dash-dotted curve, which is almost vertical line very close to the y-axis, is the boundary of the PDFI state absolute instability. The PDFI state exists in a wide region between the solid and dash-dotted curves and coexists with the SDFI state above the dotted curve. Moving right above the dotted curve $\bar{P}$ increases and $A$ decreases meaning that the relative fraction of the SDFI state also increases.

From **Fig. 3(a)**, big ratios $h/\lambda \gg 1$ (corresponding to thick films) favor the stability of the coexisting SDFI and PDFI states. Big ratios $d/\lambda > 1$ (corresponding to thick dielectric gaps) favor the appearance and stability of the PDFI state. The PE phase is absolutely stable at small ratios $h/\lambda < 1$ (corresponding to the poorly screened thin films). The pure SDFI phase is absolutely stable at very small ratios $d/\lambda \leq 0.01$ (corresponding to very thin dielectric gaps and thus high screening degrees).

From **Fig. 3(b)**, the NC state ($\Delta C < 0$) exists at room temperature in a relatively big region below the contour $\Delta C = 0$. The NC state occupies of big region of PDFI states below the dotted curve in **Fig. 3(a)** and the small region of the PE phase, as well as a significant part of the rippled SDFI+PDFI state above the dotted curve in **Fig. 3(a)**. The magnitude of $\Delta C$ reaches high negative values (less than -10) when approaching the SDFI-PDFI boundary. Formally, $\Delta C$ should tends to $-\infty$ at the boundary. Since the PDFI state coexist with the SDFI state above the dotted curve, the domains existence smears the divergency. However, the NC effect is high in the significant part of the parameter space above the SDFI boundary, where the PDFI and the SDFI states coexist. The area of the NC state region increases significantly with the increase in $d/\lambda$ and decreases with the increase in $h/\lambda$. These trends are consistent with the known statement that the NC effect can be observed in the bi-layer capacitor with the ferroelectric layer is in the "shallow-well" PE phase or in the polydomain polarized state with a small amplitude of the spontaneous polarization. In the case the NC effect is induced by the strong depolarization field effects coming from the dielectric layer and incomplete screening.

From **Fig. 3(c)**, the CIPS layer apparent capacitance $\Delta C_f$ is negative in the region $\Delta C < 0$. Since $\Delta C_f \sim \frac{1}{\Delta C}$ per Eq.(13), the negative values of $\Delta C_f$ strongly increases when approaching the lower and upper boundaries of the NC state region (where $\Delta C \to 0$) and jumps from the $-\infty$ to $+\infty$ at the boundary where $\Delta C$ changes its sign. Note that $\Delta C_f \leq -1$ in the region, and the range of parameters for which $-1 < \Delta C_f \leq 0$ is absent in the considered model. The area of the region $\Delta C_f < 0$ increases significantly with the increase in $d/\lambda$ and decreases with the increase in $h/\lambda$. These trends show that the NC state can exist in the size-induced PE phase of the CIPS film, in the "shallow" PDFI state with a small amplitude $A$ of the domain modulation, as well as in the rippled SDFI+PDFI state with small average polarization $\bar{P}$ and modulation $A$.

The dependences shown in **Fig. 3(a)-(c)** are calculated at room temperature. The same dependences calculated for higher temperatures, which are still well below the transition temperature of a bulk CIPS into the PE phase, have a qualitatively similar behavior, but the area of the NC region increases significantly with increase in $T$ (see, e.g., the right column of **Fig. D1** in **Appendix D** [21] calculated for $T = 313$ K).

The dependence of $\bar{P}$, $\Delta C$, and $\Delta C_f$ on the relative size $d/\lambda$ and temperature $T$ calculated for $h/d = 5$ are shown in **Fig. 3(d), 3(e)** and **3(f)**, respectively. Black solid and dotted curves in **Fig. 3(d)** have the same meaning as in **Fig. 3(a)**. From **Fig. 3(d)**, small ratios $\lambda/d \ll 1$ (corresponding to thin dielectric gaps and/or small screening lengths) and lower temperatures ($T < 320$ K) favor the stability of the coexisting SDFI and PDFI states, which gradually transform into the unipolar state at $\lambda/d \leq 0.01$. The ratios $\lambda/d$ bigger than 0.01 favor the appearance of the PDFI state even at temperatures lower than 250 K. The PDFI-PE boundary very slightly depends on $\lambda/d$, namely, the PE phase is absolutely stable at $T > 320$ K for $0.01 < \lambda/d < 1$.

From **Fig. 3(e)**, the NC effect ($\Delta C < 0$) exists at fixed $h/d = 5$ in a relatively wide region between two black zero contours ($\Delta C = 0$). The NC state occupies of entire region of PDFI states and the region of PE phase below 330 K, as well as the significant part of the rippled SDFI+PDFI state (i.e., the part of the region below the dotted curve in **Fig. 3(d)**). The magnitude of $\Delta C$ reaches high negative values (less than -10) when approaching the boundary of the SDFI state absolute instability. Formally, $\Delta C$ should tends to $-\infty$ at the PDFI-SDFI boundary. The area of the NC region decreases significantly when $\lambda/d$ becomes smaller than 0.1 and the temperature decreases below 320 K. The NC effect is absent above 330 K corresponding to the "deep" PE phase of the bulk CIPS.

From **Fig. 3(f)**, the CIPS layer apparent capacitance $\Delta C_f$ is negative in the region where $\Delta C < 0$, however its absolute value is not smaller than 1 ($|\Delta C_f| \geq 1$) in the region. The negative values of $\Delta C_f$ strongly increases when approaching the boundaries of the NC region (where $\Delta C \to 0$) and jumps from the $-\infty$ to $+\infty$ at the boundaries. The trends shown in **Fig. 3(d)** and **3(f)** reflect the fact that the NC effect appears in the size-induced "shallow" PE phase of the CIPS film as well as in the "shallow" polydomain and rippled states with a small polarization.

The dependences shown in **Fig. 3(d)-(f)** are calculated for $h/d = 5$. The same dependences calculated for higher $h/d$ demonstrate a similar behavior, but the area of the NC region decreases significantly with increase in $h/d$ (see, e.g., the right column of **Fig. D2** in **Appendix D** calculated for $h/d = 20$).

Note that the effects of partial or full single-domenization induced by the transverse confinement in finite-size CIPS plates should be considered under the calculations of capacitance in the coexistence region of the SDFI and PDFI phases. Under the condition $L < D_{cr}$ the unipolarity degree increases in the coexistence region, and the PDFI phase disappears for $L < \frac{D_{cr}}{2}$. For small $L$, which are of special interest for the short-channel FETs, the condition $L < \frac{D_{cr}}{2}$ can be valid for a given $h$, $d$ and $\lambda$. The PDFI phase is virtually absent in this practically important case and so the effective capacitance can reach giant negative values near the virtual SDFI-PE boundary (shown by dotted curves in **Figs. 3(a)** and **3(d)**). Thus, the dark-blue regions between the contours labeled "-10" in **Figs. 3(b)** and **3(e)**, where $\Delta C < -10$, should be considered as maximal NC effect possible for the case $L < \frac{D_{cr}}{2}$.

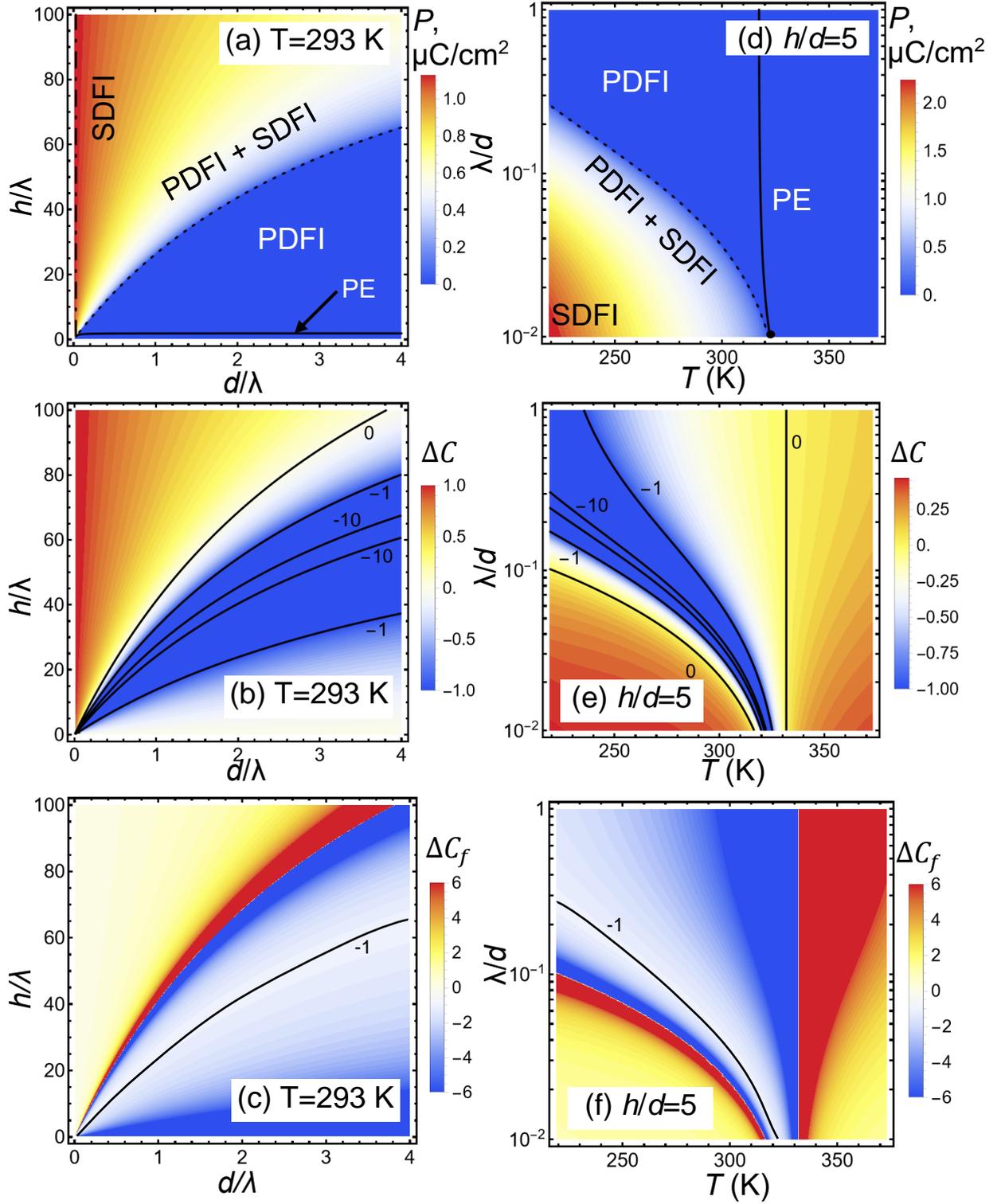

**Figure 3.** The dependences of the average spontaneous polarization in µC/cm² (**a**), relative capacitance $\Delta C$ (**b**), and the apparent capacitance of the CIPS layer $\Delta C_f$ (**c**) on the relative thickness $h/\lambda$ of the CIPS film and relative thickness $d/\lambda$ of the SiO$_2$ layer calculated for $T$=293 K. The dependences of the average spontaneous polarization in µC/cm² (**d**), relative capacitance $\Delta C$ (**e**), and the apparent capacitance of the CIPS layer $\Delta C_f$ (**f**) on the dimensionless screening length $\frac{\lambda}{d}$ in the 2D-MoS$_2$ layer and temperature $T$ calculated for the ratio $\frac{h}{d}$ =5. Misfit strain $u_m = 0.3\%$, $\varepsilon_d = 3.9$. Small black numbers near the contour lines in the parts (b), (c), (e) and (f) correspond to the values of $\Delta C$ and $\Delta C_f$, respectively.

The dependence of the interfacial electrostatic potential, $\varphi(h)$, on the relative sizes $h/\lambda$ and $d/\lambda$ calculated at room temperature and three voltages $U$ within the range (3 – 300) mV is shown in **Fig. 4(a)-(c)**. The same dependences calculated for more voltage values are shown in **Fig. D3** in **Appendix D** [21]. Distributions of the electrostatic potential across the heterostructure "CIPS film – 2D-MoS$_2$ layer – SiO$_2$ layer" calculated in the temperature range (268 – 318) K and voltage range (0 – 300) mV for $\lambda$=0.5 nm, $d$ = 4 nm and $h$= 20 nm are shown in **Fig. 4(d)-(f)**. Since all parts of the figure are calculated in the single-domain approximation (e.g., for $L < \frac{D_{cr}}{2}$), the potential is independent on the transverse coordinates x and y.

The contrast between the regions with very small positive, zero, and negative $\varphi(h)$ is clearly present for small positive $U \sim$ mV (**Fig. 4(a)**). The contrast fully correlates with the boundaries between the PE phase and the SDFI state in **Fig. 3(a)**. The contrast rapidly getting worse with the voltage increase above 10 mV (see **Fig. 4(b)**) and completely vanishes above 100 mV, however the diffuse region with negative $\varphi(h)$ remains (see **Fig. 4(c)**).

Note that the different signs of applied voltage $U$ and interfacial potential $\varphi(h)$ is the feature of the NC effect. Indeed, corresponding distributions of the electrostatic potential across the heterostructure "CIPS film – 2D-MoS$_2$ layer – SiO$_2$ dielectric layer" contain the region $\varphi(z) < 0$ for all z inside the CIPS film and inside a significant part of the dielectric layer (see **Fig. 4(d)-4(f)**). The potential $\varphi(z)$ is zero at the bottom electrode $z = 0$, becomes negative entire the CIPS film, linearly decreases and reaches maximal negative values at the 2D-MoS$_2$ interface $z = h$, then is linearly increases in the dielectric layer and changes sign to reach the positive $U$ values at the top electrode $z = h + d$. The region $\varphi(z) < 0$, which covers the CIPS film and the part of the dielectric layer, exists at all temperatures in the range (268 – 318) K and voltages in the range (3 – 300) mV. This provides the pronounced NC effect in the studied heterostructure and allows its temperature and voltage control.

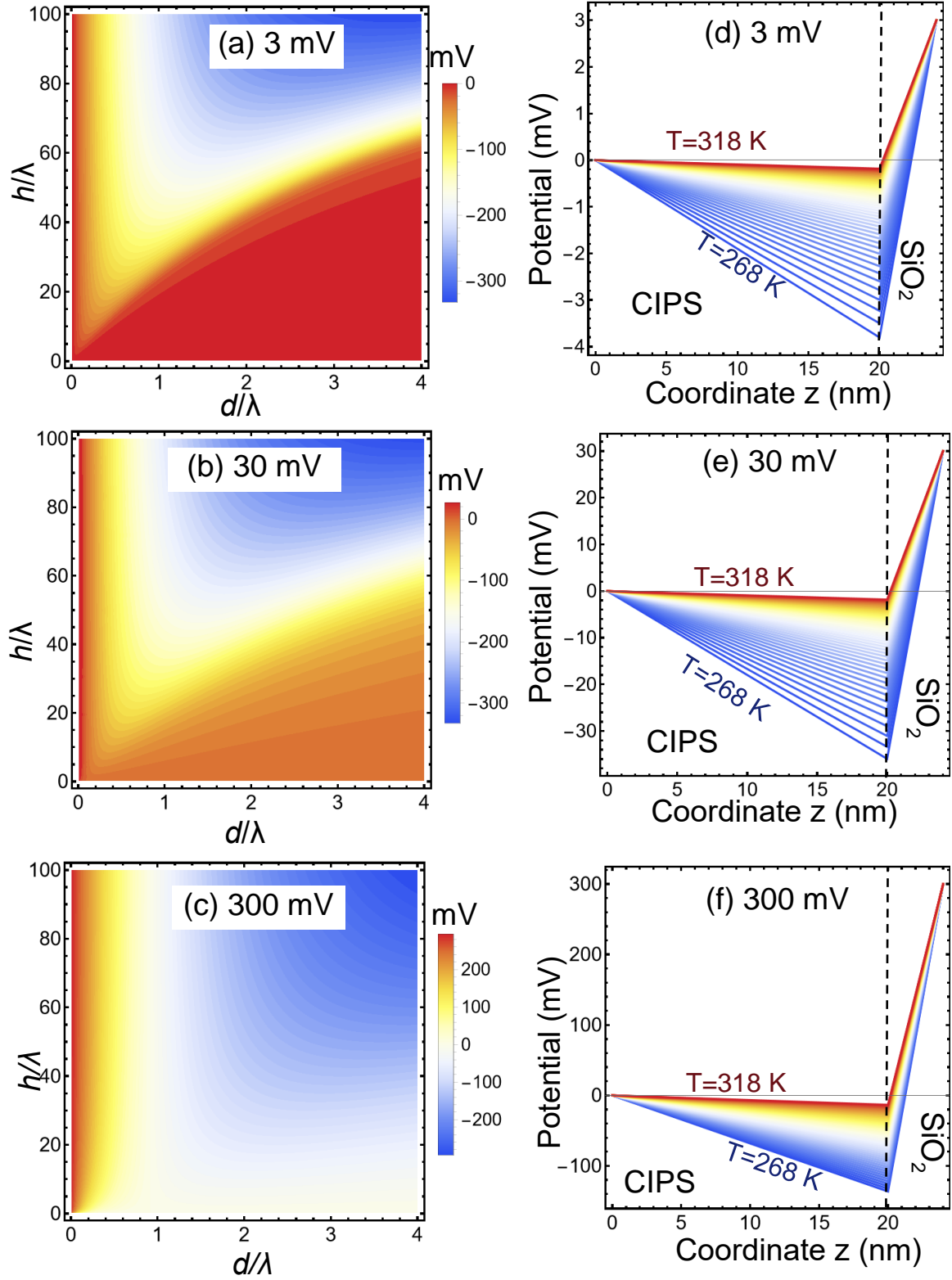

**Figure 4.** The dependence of the electrostatic potential at the interface, $\varphi(h)$, on the relative thickness $h/\lambda$ of the CIPS film and relative thickness $d/\lambda$ of the SiO$_2$ layer for the temperature $T$=293 K and different values of applied voltage, $U = 0$ **(a)**, 30 mV **(b)**, and 300 mV **(c)**. Distribution of the electrostatic potential across the heterostructure "CIPS film – 2D-MoS$_2$ single-layer – SiO$_2$ layer" calculated in the temperature range from 268 (blue curves) to 318 K (red curves) with the step of 1 K and different values of applied voltage, $U = 3$ mV **(d)**,

30 mV **(e)** and 300 mV **(f),** screening length $\lambda$=0.5 nm, $d = 4$ nm, $h$= 20 nm. Other parameters are the same as in **Fig. 3**.

## B. Results of the Finite Element Modelling

To study the influence of the domain formation on the electric polarization, electric and elastic fields, and electrostatic potential distributions beyond the harmonic approximation (7), we performed the FEM of these quantities in the studied heterostructure "CIPS film – 2D-MoS$_2$ single-layer – SiO$_2$ layer". The FEM allows to determine the period of the domain structure from the minimum of LGD free energy.

Typical distributions of the CIPS polarization and electrostatic potential inside the heterostructure calculated for CIPS thickness $h$ from 2 nm to 20 nm, fixed $d = 4$ nm and $\lambda =0.5$ nm, $U = 0$ and $T =293$ K are shown in **Fig. 5.** It is seen that the domains are absent in the 2-nm thick film, which is in the PE phase. Corresponding potential is zero (see **Fig. 5(a)**). The faint 180-degree domain stripes appear in the 2.5 nm film, being more pronounced near the top electrode ($z = 0$) and much less visible near the interface ($z = h$) due to the significant broadening of the domain walls caused by the incomplete screening by 2D-MoS$_2$ [27]. Contrary, the amplitude of potential modulation in the transverse x-direction is maximal near the interface due to the highest depolarization field in the spatial region. The modulation disappears at small distance from the interface because the depolarization field rapidly vanishes. The period of the potential modulation is equal to the period of the domain stripes, as anticipated (see **Fig. 5(b)**). The 180-degree domain stripes become much more contrast and its period significantly increases with the film thickness increase from 5 nm to 20 nm (see **Figs. 5(c)-5(f)**). At the same time the domain walls broadening also becomes stronger near the interface $z = h$ due to the strong depolarization field in the spatial region. In result the transverse modulation of the electrostatic potential becomes very pronounced near the interface; its amplitude reaches maximum at the interface (i.e., inside the ultra-thin MoS$_2$ single-layer) and gradually decreases far from the interface. Notably, that the contrast modulation of the potential exists in the SiO$_2$ layer too, but the modulation penetration depth in SiO$_2$ layer is much smaller than in the CIPS film. The potential modulation in the SiO$_2$ layer is related with the existence of the stray electric field induced by the domain stripes. However, the penetration depth of the stray electric field is significantly smaller in the SiO$_2$ layer than the penetration depth of the depolarization electric field in the CIPS layer. From Eq.(7b) the field penetration depths are defined by the parameter $\xi$ for the CIPS and by the wavenumber $k$ for the SiO$_2$. The ratio $\xi/k$ is equal to the expression $\sqrt{\left(\frac{2}{\pi}kh\right)^2 + \left(\frac{\varepsilon_d k}{\tanh(kd)} + \frac{1}{\lambda}\right)\frac{h}{\varepsilon_b}}$. For the equilibrium value of $k$ the expression is proportional to $\sqrt{\frac{\varepsilon_b}{\varepsilon_f}}$,

where $\varepsilon_f$ is the relative dielectric permittivity of a ferrielectric CIPS, which is an order of magnitude higher than its background permittivity $\varepsilon_b = 9$ (see **Appendix C** [21] for details).

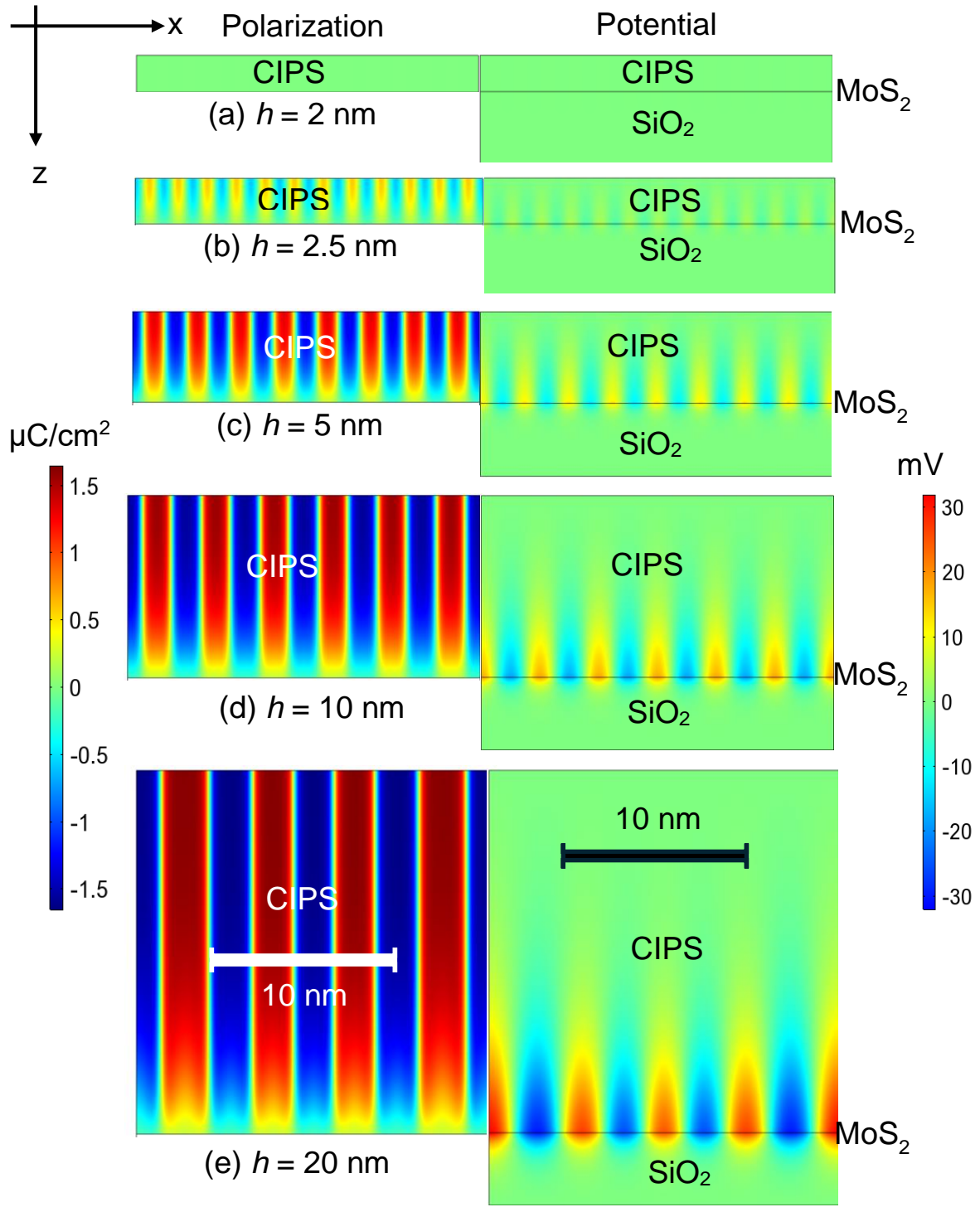

**Figure 5**. Distributions of the CIPS polarization (left column) and the electrostatic potential inside the heterostructure "CIPS film – 2D-MoS$_2$ single-layer – SiO$_2$ layer" (right column) calculated for the SiO$_2$ layer thickness $d = 4$ nm and CIPS thicknesses $h = 2$ nm (**a**), 2.5 nm (**b**), 5 nm (**c**), 10 nm (**d**) and 20 nm (**e**). Parameters: $U = 0$, $\lambda = 0.5$ nm, $\varepsilon_d = 3.9$, misfit strain $u_m = 0.3\%$, and $T = 293$ K.

Let us underline that the period of domain stripes calculated by the FEM, which increases from 1.4 nm to 4.5 nm under the film thickness increase from 2.5 nm to 20 nm, well agrees with the period calculated analytically from Eq.(7c) and shown by the blue lines in **Fig. 2(a)** and **2(b).** The agreement corroborates the validity of the analytical expressions (7) for calculations of the poly-domain states in the phase diagrams shown in **Fig. 3**.

Typical distributions of the CIPS polarization and the electrostatic potential inside the CIPS-MoS$_2$-SiO$_2$ heterostructure calculated for applied voltages $U$ from 3 mV to 300 mV, $h = 20$ nm and the same other parameters as in **Fig. 5**, are shown in **Fig. 6**. It is seen that the application of 3 mV and 30 mV practically does not change the width of "up" and "down" the domain stripes (see **Fig. 6(a)** and **6(b)**), while the application of 300 mV makes the stripes with "up" polarization direction significantly wider and corresponding modulation of the electrostatic potential becomes hardly seen (see **Fig. 6(c)**). It is seen from **Fig. D4** in **Appendix D** [21] that very thin domain stripes with "down" polarization are still present at $U = 0.5$ V, and the CIPS film becomes single domain at applied voltages about 1 V or higher.

Results shown in **Figs. 5** and **6** fully correlate with the behavior of potential contrast shown in **Fig. 4**, despite the figure is calculated in a single-domain approximation. Hence, we can consider the interfacial potential $\varphi(h)$ given by Eq.(10a) and shown in **Fig. 4**, as the average value of the electrostatic potential that can be modulated by the presence of the domain stripes.

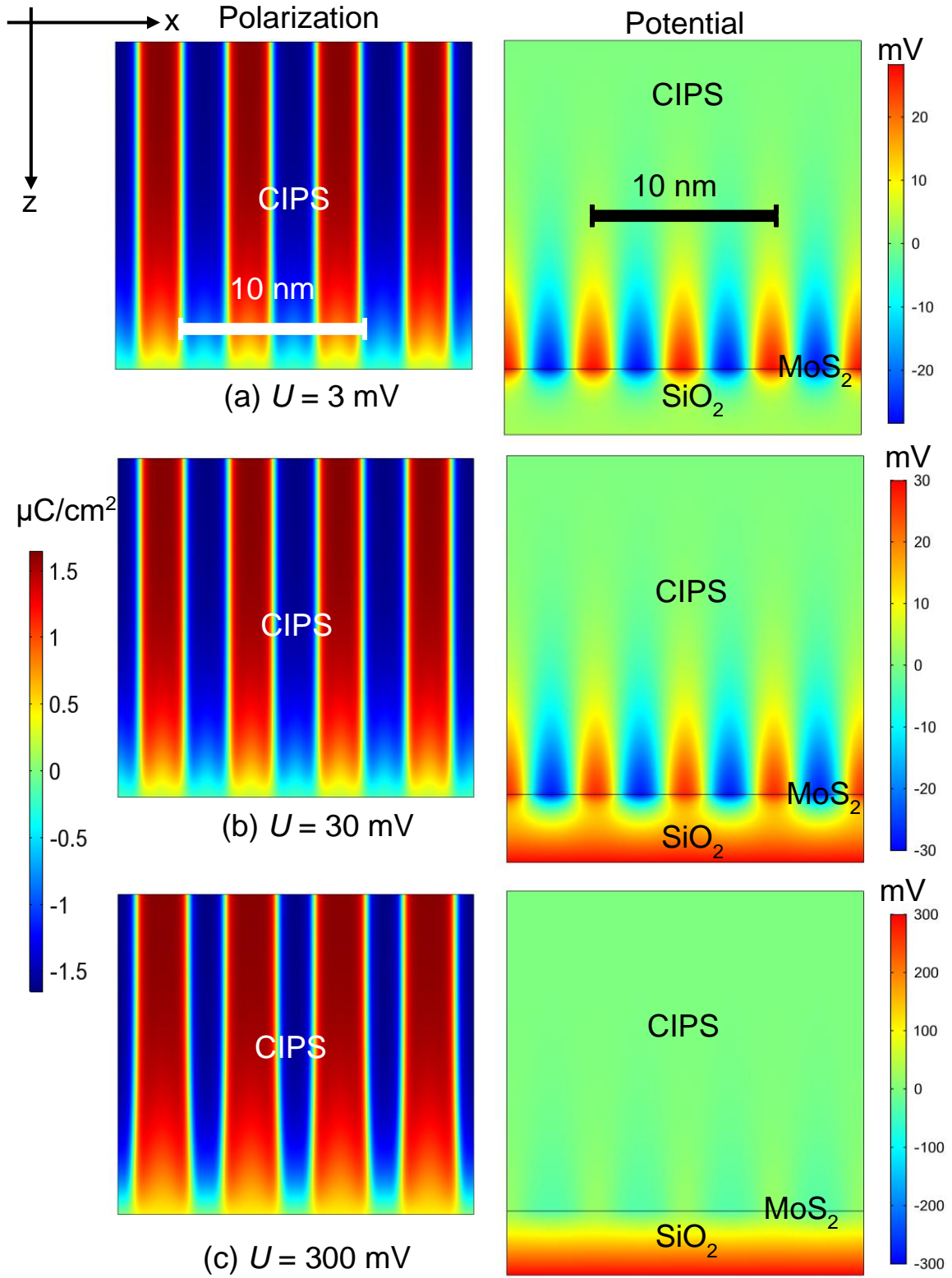

**Figure 6**. Distributions of the CIPS polarization (left column) and the electrostatic potential inside the heterostructure "CIPS film – 2D-MoS$_2$ single-layer – SiO$_2$ layer" (right column) calculated for different values of applied voltage $U = 3$ mV (**a**), 30 mV (**b**), and 300 mV (**c**). Parameters: $h = 20$ nm, $\lambda$=0.5 nm, $d = 4$ nm, $\varepsilon_d = 3.9$, $u_m = 0.3\ \%$, and $T = 293$ K.

## C. Analytical Solutions for the Channel Subthreshold Swing

The subthreshold swing $S$ shows how many times it is necessary to increase the gate voltage $V_g$ in the subthreshold region to achieve an increase in the drain current $I_d$ by an order of magnitude:

$$S = \frac{\partial V_g}{\partial (\log_{10} I_d)} = \ln 10 \frac{\partial V_g}{\partial \varphi(x,h)} \frac{\partial \varphi(x,h)}{\partial (\ln I_d)}. \tag{14}$$

For a high-quality FET with a large value of the positive gate capacitance, the limiting value of $S$ is equal to $\ln 10 \frac{k_B T}{e}$, where $e$ is the electron charge, $k_B$ is the Boltzmann constant and $T$ is the temperature in Kelvins. At room temperature $S \approx 60$ mV/dec. The importance of the subthreshold swing reduction follows from the fact that its smallest limiting value determines the minimal possible operating voltage of the transistor supply. Below we analyze the size effect of $S$ for the NC-FET shown in **Fig. 7(a)**.

Due to the possible presence of domain stripes and motion of the domain walls, the considered problem is two-dimensional: $x$ is the axis along the channel of the transistor with length $L$, and $z$ is perpendicular to the plane (see **Fig. 1(a)**). Our goal is to find the dependence of the potential $\varphi(x,z)$, as a function of the gate and source-drain voltages, $V_g$ and $V_d$, by a self-consistent solution of the Poisson and LGD equations. Next, one can find the current density in the infinitely thin 2D-MoS₂ channel ($\vec{j}(x)$) and the current ($I(x)$) from the standard expressions:

$$j_x(x) = \sigma_s(x) \mu_n E_x(x), \tag{15a}$$

$$I(x) \approx -W \sigma_s(x) \mu_n \frac{d}{dx} \varphi(x, h) = I_d. \tag{15b}$$

Here $\mu_n$ is the mobility, $\sigma_s(x)$ is the surface charge density of 2D-carriers in the MoS₂ channel and $W$ is the channel width, which is regarded equal to a lattice constant.

The condition $I(x) = I_d$ follows from the requirement of the $x$-independent $\vec{j}_n(x)$. When domain stripes exist in CIPS film there is no equipotentiality in the channel, and thus $\varphi(x,h)$ changes along $x$ (see e.g., **Figs. 5** and **6**). Since the dragging voltage $V_d$ is applied along the channel, the density $\sigma_s(x)$ is different from the equilibrium surface charge density $\sigma(x)$, which is regarded equal to $-\varepsilon_0 \frac{\varphi(x,h)}{\lambda}$ in accordance with Eq.(6). To fulfill the condition of constant $I(x)$, we should postulate that the product $\sigma_s(x) \frac{d}{dx} \varphi(x,h)$ is constant. To find the current, it is necessary to take the product of $\varphi(x,h)$ and $\sigma_s(x)$ at a certain point, for example, at the point of the virtual sourse $x_o$, where the potential has an extremum in $x$. Having found the drain current from Eq.(15), the subthreshold swing can be calculated from Eq.(14).

Since we have shown in the previous subsection that we can consider the interface potential $\varphi(h)$ given by Eq.(10a), as the average value of the electrostatic potential, the virtual source approximation is valid for the "shallow" domain stripes with a small polarization amplitude. For the equivalent scheme of NC-FETs shown in **Fig. 7(b)**, $\frac{\partial V_g}{\partial \varphi(h)} \approx \left(1 + \frac{C_d}{C_f}\right)$, and so the impact of the heterostructure capacitance on the subthreshold swing can be estimated as [20]:

$$S \approx \left(1 + \frac{C_d}{C_f}\right) S_0, \qquad S_0 = \ln 10 \frac{k_B T}{e}. \tag{16}$$

From Eqs. (13) and (16) we obtained the following expression for $S$ at room temperature:

$$S \approx \left(1 + \frac{1}{\Delta C_f}\right) 60 \frac{\text{mV}}{\text{dec}} = \left(1 + \frac{\frac{1}{2\lambda} + \frac{\varepsilon_d}{d}}{\frac{1}{2\lambda} + \frac{\varepsilon_a}{h}}\right) 60 \text{ mV/dec}. \tag{17}$$

Since $\Delta C_f$ can be negative in dependence on sizes $h$, $d$ and screening length $\lambda$ (see, e.g., **Fig. 3(c)** and **3(f)**), we can select the range of parameters, where $\Delta C_f < 0$, and thus $S$ is smaller than the Boltzmann limit in the range. The closer is $\Delta C_f$ to -1, the steeper is the subthreshold swing; and $S$ becomes much smaller the Boltzmann limit (see **Figs. 7(c)** and **7(d)**).

Since the condition $-1 < \Delta C_f < 0$ is hypothetically possible, one could obtain $S < 0$ in the case. In the case a hysteresis behavior may occur and cause the device instability issue and circuit error, impeding the application of NC-FET in logic circuits [28]. However, the range of parameters for which $-1 < \Delta C_f \leq 0$ is absent in the considered model (see e.g., **Figs. 3(c)** and **3(f)**).

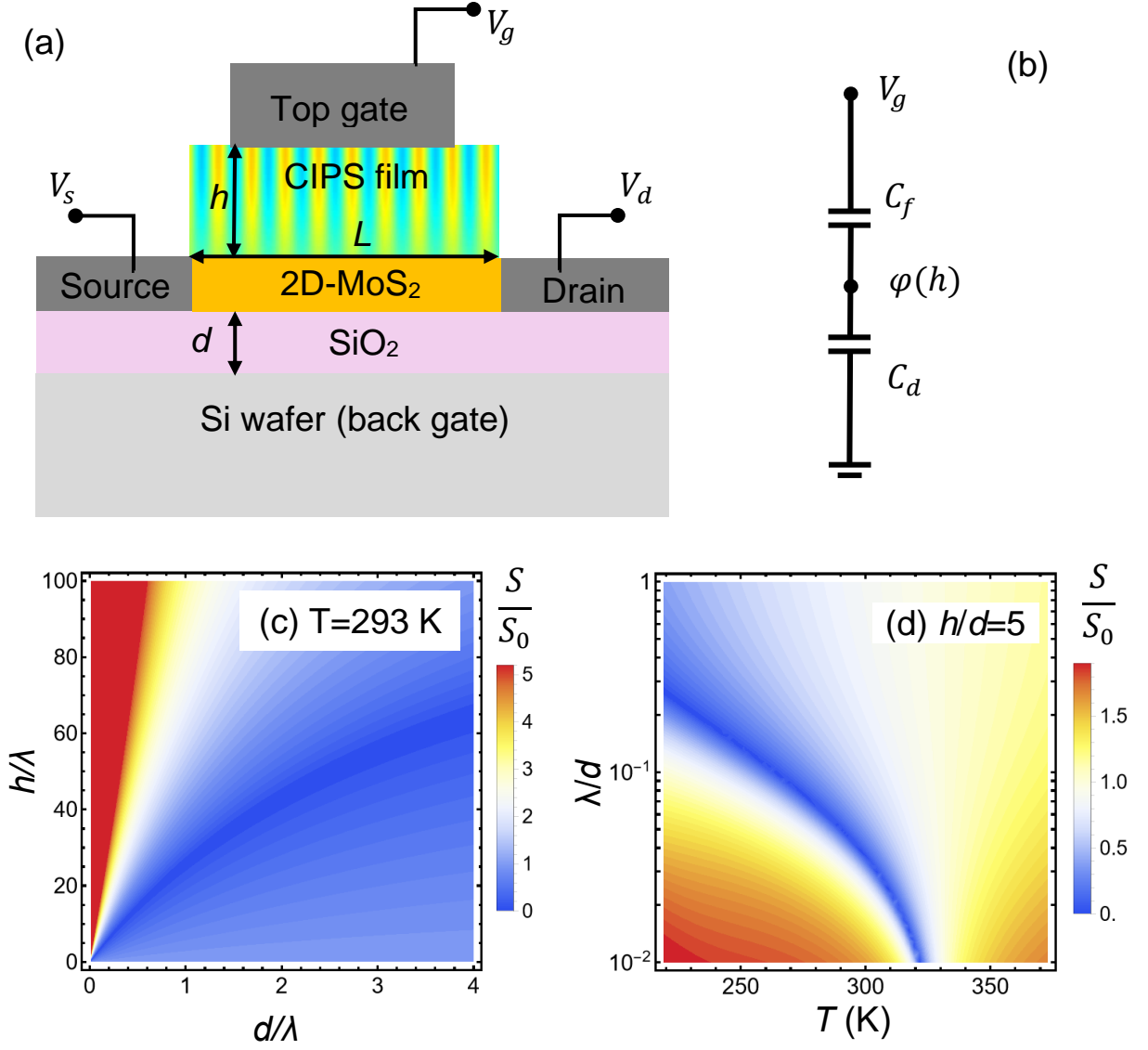

**Figure 6.** The NC-FET **(a)** and its equivalent scheme **(b)**. **(c)** The dependence of the relative subthreshold swing $S/S_0$ on the relative thickness $h/\lambda$ of the CIPS film and relative thickness $d/\lambda$ of the SiO$_2$ layer calculated at $T$=293 K. **(d)** The dependence of the $S/S_0$ on the dimensionless screening length $\frac{\lambda}{d}$ in the 2D-MoS$_2$ single-layer and temperature $T$ calculated for the ratio $\frac{h}{d}$ =5. Misfit strain $u_m = 0.3\%$, $\varepsilon_d = 3.9$.

## V. CONCLUSIONS

We perform analytical calculations in the framework of Landau-Ginzburg-Devonshire (LGD) approach and finite element modelling (FEM) of the electric and elastic fields, polarization and capacitance in the FET-type heterostructure "CIPS film – 2D-MoS$_2$ single-layer – SiO$_2$ dielectric layer". Using a direct variational method, we derived analytical expressions for the ferrielectric polarization average value, amplitude and period of the emergent domain structure, which allows to calculate the spontaneous polarization and multi-layer capacitance phase diagrams.

We derived analytical expressions, which allow to predict the thickness range of the dielectric layer and ferrielectric film for which the NC effect is the most pronounced in CIPS and the corresponding subthreshold swing becomes much less than the Boltzmann's limit (60 mV/decade at room temperature). These results explain and quantify those observed experimentally in a similar heterostructure by Wang et al. [14]. The physical origin of the subthreshold swing reduction is the NC effect induced by specific energy-degenerated poly-domain states of the spontaneous polarization, which appear in the CIPS film under incomplete screening conditions in the presence of a dielectric layer.

Obtained analytical expressions predict the dominant role of finite size effects in the multi-layer NC heterostructures and allow to predict the range of the dielectric and ferrielectric layer thickness film for which the NC effect is the most pronounced in various Van der Waals ferrielectrics. The high-performance NC state reduces the subthreshold swing of the NC-FET well below the Boltzmann's limit. Thus, the analytical results corroborated by FEM can be useful for the size and temperature control of the NC effect in the steep-slope ferrielectric FETs.

**Acknowledgements.** The work of A.N.M. and E.A.E. is supported by the DOE Software Project on "Computational Mesoscale Science and Open Software for Quantum Materials", under Award Number DE-SC0020145 as part of the Computational Materials Sciences Program of US Department of Energy, Office of Science, Basic Energy Sciences. This effort (problem statement and general analysis, S.V.K.) was supported as part of the center for 3D Ferroelectric Microelectronics (3DFeM), an Energy Frontier Research Center funded by the U.S. Department of Energy (DOE), Office of Science, Basic Energy Sciences under Award Number DE-SC0021118. Y.M.V. acknowledges support from the Horizon Europe Framework Programme (HORIZON-TMA-MSCA-SE), project № 101131229, Piezoelectricity in 2D-materials: materials, modeling, and applications (PIEZO 2D). Results were visualized in Mathematica 14.0 [29].

**Authors' contribution.** A.N.M., S.V.K. and M.V.S. generated the research idea, formulated the problem and wrote the manuscript draft. A.N.M. and E.A.E. performed analytical calculations and prepared figures. E.A.E. wrote the codes. Y.M.V. worked on the results explanation and manuscript improvement. All co-authors discussed the results.

# APPENDIX A. Calculation details

**Table AI.** LGD parameters for a CuInP$_2$S$_6$ film

| Coefficient | Units | Numerical value |
|---|---|---|
| $\varepsilon_b$ | dimensionless | 9 |
| $\alpha_T$ | C$^{-2}$·m J/K | $1.64067 \times 10^7$ |
| $T_C$ | K | 292.67 |
| $\beta$ | C$^{-4}$·m$^5$J | $3.148 \times 10^{12}$ |
| $\gamma$ | C$^{-6}$·m$^9$J | $-1.0776 \times 10^{16}$ |
| $\delta$ | C$^{-8}$·m$^{13}$J | $7.6318 \times 10^{18}$ |
| $Q_{i3}$ | C$^{-2}$·m$^4$ | $Q_{13} = 1.70136 - 0.00363\,T$, $Q_{23} = 1.13424 - 0.00242\,T$, $Q_{33} = -5.622 + 0.0105\,T$ |
| $Z_{i33}$ | C$^{-4}$·m$^8$ | $Z_{133} = -2059.65 + 0.8\,T$, $Z_{233} = -1211.26 + 0.45\,T$, $Z_{333} = 1381.37 - 12\,T$ |
| $W_{ij3}$ | C$^{-2}$·m$^4$ Pa$^{-1}$ | $W_{113} \approx W_{223} \approx W_{333} \cong -2 \times 10^{-12}$ |
| $s_{ij}$ | Pa$^{-1}$ | $s_{11} = 1.510 \times 10^{-11}$, $s_{12} = 0.183 \times 10^{-11}$ |
| $g_{33ij}$ | J m$^3$/C$^2$ | $g_{33} = 5 \times 10^{-10}$, $g_{44} = 0.2 \times 10^{-10}$ |
| $d$ | nm | 2 - 20 |
| $\lambda$ | nm | 0.1 - 5 |
| $h$ | nm | 5 - 100 |

## A2. Calculations of the free charge at the electrodes (1D case)

Let us consider a multilayer capacitor consisting of a top electrode, the CIPS film of thickness $h$, the single-layer MoS$_2$, which is treated as a 2D-semiconductor with the effective screening length $\lambda$, a dielectric SiO$_2$ layer of thickness $d$, and the bottom electrode. The multilayer capacitor is shown in **Fig. 1**.

The connection between the electric displacement $\vec{D}$ and the electric field, $\vec{E} = -\nabla\varphi$, in the ferrielectric ($f$) CIPS and the dielectric($d$) SiO$_2$ layers are:

$$\vec{D}^f = \varepsilon_0 \varepsilon_b \vec{E}^f + \vec{P}, \quad \vec{D}^d = \varepsilon_0 \varepsilon_d \vec{E}^l, \tag{A.1}$$

where $\varepsilon_0$ is a universal dielectric constant, $\varepsilon_b$ is a relative background permittivity of the ferrielectric, $\varepsilon_d$ is relative dielectric permittivity of the dielectric layer. Since div $\vec{D} = 0$ in the absence of free space charges. In the case when all variables depend on the $z$-coordinate and the CIPS polarization $\vec{P}$ is directed along the polar axis $z$, $\vec{P} = (0, 0, P)$, we obtain the following electrostatic equations in the CIPS and dielectric layers:

$$\varepsilon_0 \varepsilon_b \frac{\partial^2 \varphi_f}{\partial z^2} = \frac{\partial P}{\partial z}, \quad \varepsilon_0 \varepsilon_s \frac{\partial^2 \varphi_d(z)}{\partial z^2} = 0. \tag{A.2}$$

The tangential components of the electric field are homogenous at the interfaces, and the normal components of the displacement differ by the value of the surface charge density $\sigma$ at the CIPS-2D-MoS$_2$-SiO$_2$ interface $z = h$. Taking this into account, the boundary conditions are as follows:

$$\varphi_d(h+d) = U, \tag{A.2a}$$
$$\varphi_f(h) = \varphi_d(h), \tag{A.2b}$$
$$D_z^d(h) - D_z^f(h) = \sigma, \tag{A.2c}$$
$$\varphi_f(0) = 0. \tag{A.2d}$$

Hereinafter $\sigma$ is the charge density of the 2D-MoS$_2$ channel.

For analytical calculations let us consider the case when the CIPS film can be in the polarized single-domain or depolarized paraelectric states. The case of the domain formation in the film will be considered numerically.

Using the superposition principle, we can present the general solution of Eqs.(A.1) in the following form:

$$\varphi_d(z) = -\frac{D_d(z-h-d)}{\varepsilon_0 \varepsilon_d} + U. \tag{A.3a}$$

$$\varphi_f(z) = -\frac{D_f z}{\varepsilon_0 \varepsilon_b} + \frac{1}{\varepsilon_0 \varepsilon_b} \int_0^z P(\tilde{z}) d\tilde{z}, \tag{A.3b}$$

Here the conditions (A.2a) and (A.2d) are already satisfied. After the substitution of the expressions (A.3) into the rest of the boundary conditions (A.2) and using expressions (A.1), one could get the following system of equations for the unknown displacements $D_f$ and $D_d$:

$$\begin{cases} U + \frac{h}{\varepsilon_0 \varepsilon_b} D_f - \frac{\int_0^h P(\tilde{z}) d\tilde{z}}{\varepsilon_0 \varepsilon_b} + \frac{d}{\varepsilon_0 \varepsilon_d} D_d = 0, \\ D_f + \sigma - D_d = 0. \end{cases} \tag{A.4}$$

The elementary transformations of Eq.(A.4) lead to the solution for the electric displacement, field and potential inside the CIPS are:

$$D_f = -\frac{d\sigma \varepsilon_b - \varepsilon_d \int_0^h P(\tilde{z}) d\tilde{z} + U \varepsilon_0 \varepsilon_b \varepsilon_d}{d\varepsilon_b + h\varepsilon_d}, \tag{A.5}$$

$$E_f = -\frac{P(z)}{\varepsilon_0 \varepsilon_b} + \frac{\varepsilon_d}{\varepsilon_0 \varepsilon_b (d\varepsilon_b + h\varepsilon_d)} \int_0^h P(\tilde{z}) d\tilde{z} - \frac{d\sigma}{(d\varepsilon_b + h\varepsilon_d)\varepsilon_0} - \frac{U \varepsilon_d}{d\varepsilon_b + h\varepsilon_d}, \tag{A.6}$$

$$\varphi_f(z) = \frac{1}{\varepsilon_0 \varepsilon_b} \int_0^z P(\tilde{z}) d\tilde{z} - z \left\{ \frac{\varepsilon_d}{(d\varepsilon_b + h\varepsilon_d)\varepsilon_0 \varepsilon_b} \int_0^h P(\tilde{z}) d\tilde{z} - \frac{d\sigma}{(d\varepsilon_b + h\varepsilon_d)\varepsilon_0} - \frac{U \varepsilon_d}{d\varepsilon_b + h\varepsilon_d} \right\}. \tag{A.7}$$

For the case of constant polarization, $P(\tilde{z}) \equiv P$, the field inside the CIPS is

$$E_f = -\frac{dP}{(d\varepsilon_b + h\varepsilon_d)\varepsilon_0} - \frac{d\sigma}{(d\varepsilon_b + h\varepsilon_d)\varepsilon_0} - \frac{U \varepsilon_d}{d\varepsilon_b + h\varepsilon_d}. \tag{A.8}$$

The electric displacement, field and potential inside the SiO$_2$ layer are

$$D_d = \frac{(h\sigma + \int_0^h P(\tilde{z}) d\tilde{z} - U\varepsilon_0 \varepsilon_b)\varepsilon_d}{d\varepsilon_b + h\varepsilon_d}, \tag{A.9}$$

$$E_d = \frac{h\sigma + \int_0^h P(\tilde{z}) d\tilde{z} - U\varepsilon_0 \varepsilon_b}{(d\varepsilon_b + h\varepsilon_d)\varepsilon_0}, \tag{A.10}$$

$$\varphi_d(z) = U - (z - d - h)\frac{h\sigma + \int_0^h P(\tilde{z}) d\tilde{z} - U\varepsilon_0 \varepsilon_b}{(d\varepsilon_b + h\varepsilon_d)\varepsilon_0}. \tag{A.11}$$

The potential at the CIPS-2D-MoS$_2$-SiO$_2$ interface has the form:

$$\varphi_f(h) = \frac{h\, d(\bar{P}+\sigma)}{(d\varepsilon_b + h\varepsilon_d)\varepsilon_0} + \frac{U\, h\, \varepsilon_d}{d\varepsilon_b + h\varepsilon_d}. \tag{A.12}$$

Charges at upper and lower electrodes, $Q_1$ and $Q_2$, are equal to

$$Q_1 = -D_z^d(h+d) = -\frac{h\sigma\varepsilon_d}{d\varepsilon_b + h\varepsilon_d} - \frac{h\bar{P}\varepsilon_d}{d\varepsilon_b + h\varepsilon_d} + \frac{U\varepsilon_0\varepsilon_b\varepsilon_d}{d\varepsilon_b + h\varepsilon_d} \tag{A.13}$$

$$Q_2 = +D_z^f(0) = -\frac{d\sigma\varepsilon_b}{d\varepsilon_b + h\varepsilon_d} + \frac{h\bar{P}\varepsilon_d}{d\varepsilon_b + h\varepsilon_d} - \frac{U\varepsilon_0\varepsilon_b\varepsilon_d}{d\varepsilon_b + h\varepsilon_d} \tag{A.14}$$

The discharge $Q$ is equal to:

$$Q = \frac{Q_1 - Q_2}{2} = \frac{-h\varepsilon_d + d\varepsilon_b}{2(d\varepsilon_b + h\varepsilon_d)}\sigma - \frac{h\varepsilon_d}{d\varepsilon_b + h\varepsilon_d}\bar{P} + \frac{\varepsilon_0\varepsilon_b\varepsilon_d}{d\varepsilon_b + h\varepsilon_d}U \tag{A.15}$$

### A3. Analytical calculations of the negative capacitance effect

The charge of the reference SiO$_2$ capacitor is $Q_d = C_d U$, where the reference capacitance is $C_d = \frac{\varepsilon_0 \varepsilon_d}{d}$. The difference of the effective and reference capacitance is given by expression:

$$\Delta C = C_d - \frac{dQ}{dU} = \frac{\varepsilon_0 \varepsilon_d}{d} + \frac{h\varepsilon_d - d\varepsilon_b}{d\varepsilon_b + h\varepsilon_d}\frac{d\sigma}{dU} + \frac{h\varepsilon_d}{d\varepsilon_b + h\varepsilon_d}\frac{d\bar{P}}{dU} - \frac{\varepsilon_0\varepsilon_b\varepsilon_d}{d\varepsilon_b + h\varepsilon_d}. \tag{A.16}$$

The NC effect corresponds to the condition $\Delta C < 0$. The magnitude of $\bar{P}$ can be estimated from the equation:

$$\Gamma\frac{d}{dt}\bar{P} + \tilde{\alpha}(T)\bar{P} + \tilde{\beta}\bar{P}^3 + \tilde{\gamma}\bar{P}^5 + \tilde{\delta}\bar{P}^7 = -\frac{d}{\varepsilon_d h + \varepsilon_b d}\frac{\bar{P}+\sigma}{\varepsilon_0} - \frac{\varepsilon_d}{\varepsilon_d h + \varepsilon_b d}U, \tag{A.17}$$

where $\sigma$ is the surface charge density in the 2D-MoS$_2$ channel.

Under the condition of negligibly small contribution of the nonlinear polarization powers in Eq.(A.17), the terms $\tilde{\beta}\bar{P}^3 + \tilde{\gamma}\bar{P}^5 + \tilde{\delta}\bar{P}^7$ can be omitted, and the derivative $\frac{d\bar{P}}{dU}$ can be estimated as:

$$\frac{d\bar{P}}{dU} \approx -\frac{\varepsilon_d}{\varepsilon_d h + \varepsilon_b d}\left(1 + \frac{d}{\varepsilon_0}\frac{d\sigma}{dU}\right)\left[\tilde{\alpha} + \frac{d}{\varepsilon_0(\varepsilon_d h + \varepsilon_b d)}\right]^{-1}. \tag{A.18}$$

Under the validity of Eq.(A.18), the NC effect can be reached under the conditions

$$\frac{h\varepsilon_d - d\varepsilon_b}{d\varepsilon_b + h\varepsilon_d}\frac{d\sigma}{dU} + \frac{h\varepsilon_d}{d\varepsilon_b + h\varepsilon_d}\frac{d\bar{P}}{dU} - \frac{\varepsilon_0\varepsilon_b\varepsilon_d}{d\varepsilon_b + h\varepsilon_d} > -\frac{\varepsilon_0\varepsilon_d}{d}, \qquad \tilde{\alpha} + \frac{d}{\varepsilon_0(\varepsilon_d h + \varepsilon_b d)} > 0. \tag{A.19}$$

Here the dependence $\sigma[\phi]$ is determined by the carrier statistics in the ultra-thin 2D-MoS$_2$ single-layer.

### APPENDIX B. Linearized surface charge dependence on the electric potential

The solutions for polarization and internal electric field could be essentially simplified if one could approximate the dependence of the surface charge at $z = h$ vs. the local potential $\varphi(h)$ in the following form

$$\sigma = -\varepsilon_0 \frac{\varphi(h)}{\lambda}. \tag{B.1}$$

Based on the approximation (B.1) the 1D-solution for the electric displacement, field and potential inside the CIPS could written as:

$$D_f = \frac{-\varepsilon_0 U}{\frac{d\,h}{\lambda\varepsilon_d\varepsilon_b}+\frac{d}{\varepsilon_d}+\frac{h}{\varepsilon_b}} + \frac{\left(\frac{d}{\lambda\varepsilon_d}+1\right)\frac{h}{\varepsilon_b}}{\frac{d\,h}{\lambda\varepsilon_d\varepsilon_b}+\frac{d}{\varepsilon_d}+\frac{h}{\varepsilon_b}}\bar{P}, \tag{B.2}$$

$$E_f = -\frac{P(z)}{\varepsilon_0\varepsilon_b} + \frac{\left(\frac{d}{\lambda\varepsilon_d}+1\right)\frac{h}{\varepsilon_b}}{\frac{d\,h}{\lambda\varepsilon_d\varepsilon_b}+\frac{d}{\varepsilon_d}+\frac{h}{\varepsilon_b}}\frac{\bar{P}}{\varepsilon_0\varepsilon_b} - \frac{U/\varepsilon_b}{\frac{d\,h}{\lambda\varepsilon_d\varepsilon_b}+\frac{d}{\varepsilon_d}+\frac{h}{\varepsilon_b}}, \tag{B.3}$$

$$\varphi_f(z) = \frac{1}{\varepsilon_0\varepsilon_b}\int_0^z P(\tilde{z})d\tilde{z} - z\left\{\frac{\left(\frac{d}{\lambda\varepsilon_d}+1\right)\frac{h}{\varepsilon_b}}{\frac{d\,h}{\lambda\varepsilon_d\varepsilon_b}+\frac{d}{\varepsilon_d}+\frac{h}{\varepsilon_b}}\frac{\bar{P}}{\varepsilon_0\varepsilon_b} - \frac{U/\varepsilon_b}{\frac{d\,h}{\lambda\varepsilon_d\varepsilon_b}+\frac{d}{\varepsilon_d}+\frac{h}{\varepsilon_b}}\right\}. \tag{B.4}$$

For the case of constant polarization, $P(\tilde{z}) \equiv \bar{P}$, the field inside ferroelectric is

$$E_f = -\frac{\frac{d}{\varepsilon_d}}{\frac{d\,h}{\lambda\varepsilon_d\varepsilon_b}+\frac{d}{\varepsilon_d}+\frac{h}{\varepsilon_b}}\frac{\bar{P}}{\varepsilon_0\varepsilon_b} - \frac{U/\varepsilon_b}{\frac{d\,h}{\lambda\varepsilon_d\varepsilon_b}+\frac{d}{\varepsilon_d}+\frac{h}{\varepsilon_b}}. \tag{B.5}$$

The electric displacement, field and potential inside the dielectric layer are

$$D_d = \frac{-\varepsilon_0 U\left\{1+\frac{h}{\lambda\varepsilon_b}\right\}}{\frac{d\,h}{\lambda\varepsilon_d\varepsilon_b}+\frac{d}{\varepsilon_d}+\frac{h}{\varepsilon_b}} + \frac{\frac{h}{\varepsilon_b}}{\frac{d\,h}{\lambda\varepsilon_d\varepsilon_b}+\frac{d}{\varepsilon_d}+\frac{h}{\varepsilon_b}}\bar{P}, \tag{B.6}$$

$$E_d = \frac{-\frac{U}{\varepsilon_d}\left\{1+\frac{h}{\lambda\varepsilon_b}\right\}}{\frac{d\,h}{\lambda\varepsilon_d\varepsilon_b}+\frac{d}{\varepsilon_d}+\frac{h}{\varepsilon_b}} + \frac{\frac{h}{\varepsilon_b\varepsilon_0\varepsilon_d}\bar{P}}{\frac{d\,h}{\lambda\varepsilon_d\varepsilon_b}+\frac{d}{\varepsilon_d}+\frac{h}{\varepsilon_b}}, \tag{B.7}$$

$$\varphi_d(z) = U - (z - d - h)\left\{\frac{-\frac{U}{\varepsilon_d}\left\{1+\frac{h}{\lambda\varepsilon_b}\right\}}{\frac{d\,h}{\lambda\varepsilon_d\varepsilon_b}+\frac{d}{\varepsilon_d}+\frac{h}{\varepsilon_b}} + \frac{\frac{h}{\varepsilon_b\varepsilon_0\varepsilon_d}\bar{P}}{\frac{d\,h}{\lambda\varepsilon_d\varepsilon_b}+\frac{d}{\varepsilon_d}+\frac{h}{\varepsilon_b}}\right\}. \tag{B.8}$$

The potential at the interface CIPS-2D-MoS$_2$-SiO$_2$ has the form:

$$\varphi_f(h) = \frac{\frac{d\,h}{\varepsilon_d\varepsilon_b\varepsilon_0}\bar{P}}{\frac{d\,h}{\lambda\varepsilon_d\varepsilon_b}+\frac{d}{\varepsilon_d}+\frac{h}{\varepsilon_b}} + \frac{U\frac{h}{\varepsilon_b}}{\frac{d\,h}{\lambda\varepsilon_d\varepsilon_b}+\frac{d}{\varepsilon_d}+\frac{h}{\varepsilon_b}}. \tag{B.9}$$

Charges at upper and lower electrodes, $Q_1$ and $Q_2$, are equal to

$$Q_1 = -D_z^d(h+d) = \frac{\varepsilon_0 U\left\{1+\frac{h}{\lambda\varepsilon_b}\right\}}{\frac{d\,h}{\lambda\varepsilon_d\varepsilon_b}+\frac{d}{\varepsilon_d}+\frac{h}{\varepsilon_b}} - \frac{\frac{h}{\varepsilon_b}}{\frac{d\,h}{\lambda\varepsilon_d\varepsilon_b}+\frac{d}{\varepsilon_d}+\frac{h}{\varepsilon_b}}\bar{P}, \tag{B.10a}$$

$$Q_2 = +D_z^f(0) = \frac{-\varepsilon_0 U}{\frac{d\,h}{\lambda\varepsilon_d\varepsilon_b}+\frac{d}{\varepsilon_d}+\frac{h}{\varepsilon_b}} + \frac{\left(\frac{d}{\lambda\varepsilon_d}+1\right)\frac{h}{\varepsilon_b}}{\frac{d\,h}{\lambda\varepsilon_d\varepsilon_b}+\frac{d}{\varepsilon_d}+\frac{h}{\varepsilon_b}}\bar{P}. \tag{B.10b}$$

The "discharge" $Q$ is equal to:

$$Q = \frac{Q_1 - Q_2}{2} = -\frac{\left(1+\frac{d}{2\lambda\varepsilon_d}\right)\frac{h}{\varepsilon_b}}{\frac{d\,h}{\lambda\varepsilon_d\varepsilon_b}+\frac{d}{\varepsilon_d}+\frac{h}{\varepsilon_b}}\bar{P} + \varepsilon_0\frac{1+\frac{h}{2\lambda\varepsilon_b}}{\frac{d\,h}{\lambda\varepsilon_d\varepsilon_b}+\frac{d}{\varepsilon_d}+\frac{h}{\varepsilon_b}}U \tag{B.11}$$

As a next step, we use the average polarization dependence on the internal field inside the CIPS layer, namely

$$\tilde{\alpha}\bar{P} + \tilde{\beta}\bar{P}^3 + \tilde{\gamma}\bar{P}^5 + \tilde{\delta}\bar{P}^7 = E_f. \tag{B.12a}$$

Or in the evident form, taking into account Eq.(B.5)

$$\left(\tilde{\alpha} + \frac{\frac{d}{\varepsilon_d}}{\frac{d\,h}{\lambda\,\varepsilon_d\,\varepsilon_b} + \frac{d}{\varepsilon_d} + \frac{h}{\varepsilon_b}}\frac{1}{\varepsilon_0\varepsilon_b}\right)\bar{P} + \tilde{\beta}\bar{P}^3 + \tilde{\gamma}\bar{P}^5 + \tilde{\delta}\bar{P}^7 = -\frac{U/\varepsilon_b}{\frac{d\,h}{\lambda\,\varepsilon_d\,\varepsilon_b} + \frac{d}{\varepsilon_d} + \frac{h}{\varepsilon_b}}. \tag{B.12b}$$

Linearized solution of Eq.(12b) is

$$\bar{P} \approx \bar{P}_0 + \delta P, \tag{B.13a}$$

where $\bar{P}_0$ and $\delta P$ satisfy the following equations:

$$\left(\tilde{\alpha} + \frac{\frac{d}{\varepsilon_d}}{\frac{d\,h}{\lambda\,\varepsilon_d\,\varepsilon_b} + \frac{d}{\varepsilon_d} + \frac{h}{\varepsilon_b}}\frac{1}{\varepsilon_0\varepsilon_b}\right)\bar{P}_0 + \tilde{\beta}\bar{P}_0^3 + \tilde{\gamma}\bar{P}_0^5 + \tilde{\delta}\bar{P}_0^7 = 0, \tag{B.13b}$$

$$\delta P = -\frac{\frac{U}{\varepsilon_b}}{\frac{d\,h}{\lambda\,\varepsilon_d\,\varepsilon_b} + \frac{d}{\varepsilon_d} + \frac{h}{\varepsilon_b}}\left(\tilde{\alpha} + \frac{\frac{d}{\varepsilon_d}}{\frac{d\,h}{\lambda\,\varepsilon_d\,\varepsilon_b} + \frac{d}{\varepsilon_d} + \frac{h}{\varepsilon_b}}\frac{1}{\varepsilon_0\varepsilon_b} + 3\beta\bar{P}_0^2 + 5\gamma\bar{P}_0^4 + 7\delta\bar{P}_0^6\right)^{-1} = \frac{\varepsilon_0 U}{\left(\frac{d\,h}{\lambda\,\varepsilon_d\,\varepsilon_b} + \frac{d}{\varepsilon_d} + \frac{h}{\varepsilon_b}\right)\varepsilon_0\varepsilon_b\alpha_0 + \frac{d}{\varepsilon_d}}. \tag{B.13c}$$

Here the renormalized coefficient $\alpha_0$ is introduced as:

$$\alpha_0 = \tilde{\alpha} + 3\tilde{\beta}\bar{P}_0^2 + 5\tilde{\gamma}\bar{P}_0^4 + 7\tilde{\delta}\bar{P}_0^6. \tag{B.13d}$$

Note that the effective dielectric stiffness can be derived from (B.12b) as:

$$\alpha_{eff} = \tilde{\alpha} + \frac{\frac{d}{\varepsilon_d}}{\frac{d\,h}{\lambda\,\varepsilon_d\,\varepsilon_b} + \frac{d}{\varepsilon_d} + \frac{h}{\varepsilon_b}}\frac{1}{\varepsilon_0\varepsilon_b} + 3\tilde{\beta}\bar{P}_0^2 + 5\tilde{\gamma}\bar{P}_0^4 + 7\tilde{\delta}\bar{P}_0^6 = \alpha_0 + \frac{\frac{d}{\varepsilon_0\varepsilon_b\varepsilon_d}}{\frac{d\,h}{\lambda\,\varepsilon_d\,\varepsilon_b} + \frac{d}{\varepsilon_d} + \frac{h}{\varepsilon_b}} \tag{B.14}$$

It is easy to show that the stiffness is always positive, $\alpha_{eff} \geq 0$, in the stable state of the system, while $\alpha_0$ can change its sign.

Below we consider the effective capacitance of the system, which is defined as:

$$C_{eff} = \frac{dQ}{dU}. \tag{B.15a}$$

The reference capacitance is

$$C_d = \frac{\varepsilon_0\varepsilon_d}{d}. \tag{B.15b}$$

After the simple expression for the flat capacitor filled with media with permittivity $\varepsilon_d$. The difference of the effective and reference capacitance is given by expression:

$$\Delta C = C_d - C_{eff} = \frac{\varepsilon_0\varepsilon_d}{d} - \left\{\varepsilon_0\frac{1+\frac{h}{2\lambda\,\varepsilon_b}}{\frac{d\,h}{\lambda\,\varepsilon_d\,\varepsilon_b} + \frac{d}{\varepsilon_d} + \frac{h}{\varepsilon_b}} - \frac{\left(1+\frac{d}{2\lambda\,\varepsilon_d}\right)\frac{h}{\varepsilon_b}}{\frac{d\,h}{\lambda\,\varepsilon_d\,\varepsilon_b} + \frac{d}{\varepsilon_d} + \frac{h}{\varepsilon_b}}\frac{d\bar{P}}{dU}\right\} \tag{B.16a}$$

Polarization derivative could be derived from (B.13c) as

$$\frac{d\bar{P}}{dU} \approx \frac{d}{dU}\delta P = -\frac{\varepsilon_0}{\left(\frac{d\,h}{\lambda\,\varepsilon_d\,\varepsilon_b} + \frac{d}{\varepsilon_d} + \frac{h}{\varepsilon_b}\right)\varepsilon_0\varepsilon_b\alpha_0 + \frac{d}{\varepsilon_d}}. \tag{B.16b}$$

The evident form of the effective capacitance of the multilayer is

$$C_{eff} = \frac{\varepsilon_0}{\frac{d\,h}{\lambda\,\varepsilon_d\,\varepsilon_b} + \frac{d}{\varepsilon_d} + \frac{h}{\varepsilon_b}}\left\{\frac{\left(1+\frac{d}{2\lambda\,\varepsilon_d}\right)\frac{h}{\varepsilon_b}}{\left(\frac{d\,h}{\lambda\,\varepsilon_d\,\varepsilon_b} + \frac{d}{\varepsilon_d} + \frac{h}{\varepsilon_b}\right)\varepsilon_0\varepsilon_b\alpha_0 + \frac{d}{\varepsilon_d}} + 1 + \frac{h}{2\lambda\,\varepsilon_b}\right\} = \frac{\varepsilon_0\left(1+\varepsilon_0\varepsilon_b\alpha_0 + \frac{h}{2\lambda\,\varepsilon_b}\varepsilon_0\varepsilon_b\alpha_0\right)}{\left(\frac{d\,h}{\lambda\,\varepsilon_d\,\varepsilon_b} + \frac{d}{\varepsilon_d} + \frac{h}{\varepsilon_b}\right)\varepsilon_0\varepsilon_b\alpha_0 + \frac{d}{\varepsilon_d}} = \frac{\varepsilon_0\varepsilon_d}{d}\frac{\frac{\varepsilon_a}{h} + \frac{1}{2\lambda}}{\frac{1}{\lambda} + \frac{\varepsilon_a}{h} + \frac{\varepsilon_d}{d}}. \tag{B.17a}$$

Here we introduced the apparent dielectric permittivity of the CIPS layer

$$\varepsilon_a = \frac{1}{\varepsilon_0 \alpha_0} + \varepsilon_b. \tag{B.17b}$$

Eventually, the effect of the NC corresponds to the $\Delta C < 0$, where

$$\Delta C = C_d - C_{eff} = \frac{\varepsilon_0 \varepsilon_d}{d}\left[1 - \frac{\frac{\varepsilon_a}{h} + \frac{1}{2\lambda}}{\frac{1}{\lambda} + \frac{\varepsilon_a}{h} + \frac{\varepsilon_d}{d}}\right] = \frac{\varepsilon_0 \varepsilon_d}{d}\left[\frac{\frac{1}{2\lambda} + \frac{\varepsilon_d}{d}}{\frac{1}{\lambda} + \frac{\varepsilon_a}{h} + \frac{\varepsilon_d}{d}}\right]. \tag{B.18}$$

We can estimate the apparent capacitance of the "CIPS +MoS$_2$" layer $C_f$ as the solution of equation $\frac{1}{C_f} = \frac{1}{C_{eff}} - \frac{1}{C_d}$, and introduce its relative value as

$$\Delta C_f = \frac{C_f}{C_d} = \frac{C_{eff}}{C_d - C_{eff}} = 1 - \frac{1}{\Delta C} = -\frac{\frac{1}{2\lambda} + \frac{\varepsilon_a}{h}}{\frac{1}{2\lambda} + \frac{\varepsilon_d}{d}}. \tag{B.19}$$

## APPENDIX C. The derivation of the transition into the poly-domain state
### C.1. The derivation of the characteristic equation

The linearized system of equations for polarization and electric potential in the considered heterostructure has the following form:

$$\tilde{\alpha} P_3 - g_{33} \frac{\partial^2 P_3}{\partial z^2} - g_{44}\left(\frac{\partial^2 P_3}{\partial x^2} + \frac{\partial^2 P_3}{\partial y^2}\right) = -\frac{\partial}{\partial z}\varphi_f, \tag{C.1a}$$

$$\left(\frac{\partial^2}{\partial x^2} + \frac{\partial^2}{\partial y^2} + \frac{\partial^2}{\partial z^2}\right)\varphi_f = \frac{1}{\varepsilon_0 \varepsilon_b}\frac{\partial P_3}{\partial z}, \tag{C.1b}$$

$$\left(\frac{\partial^2}{\partial x^2} + \frac{\partial^2}{\partial y^2} + \frac{\partial^2}{\partial z^2}\right)\varphi_d = 0. \tag{C.1c}$$

For the domain onset one can consider the harmonic-form periodic fluctuations of the electric polarization, potential and surface charge:

$$P_3 = P_k(z)\exp(ikx), \quad \varphi_f = \phi_k^{(f)}(z)\exp(ikx), \tag{C.2a}$$

$$\sigma = \sigma_k \exp(ikx) \approx -\varepsilon_0 \frac{\phi_k^{(f)}(h)}{\lambda}\exp(ikx), \quad \varphi_d = \phi_k^{(d)}(z)\exp(ikx). \tag{C.2b}$$

The equations for the amplitudes follow from Eqs.(C.1):

$$(\tilde{\alpha} + g_{44}k^2)P_k - g_{33}\frac{\partial^2 P_k}{\partial z^2} = -\frac{\partial}{\partial z}\phi_k^{(f)}, \tag{C.3a}$$

$$\frac{\partial^2 \phi_k^{(f)}}{\partial z^2} - k^2 \phi_k^{(f)} = \frac{1}{\varepsilon_0 \varepsilon_b}\frac{\partial P_k}{\partial z}, \tag{C.3b}$$

$$\frac{\partial^2 \phi_k^{(out)}}{\partial z^2} - k^2 \phi_k^{(d)} = 0. \tag{C.3c}$$

The boundary conditions are:

$$\left(\frac{\partial P_k}{\partial z}\right)\bigg|_{z=0,h} = 0, \tag{C.4a}$$

$$\left(\phi_k^{(d)}\right)\bigg|_{z=h+d} = 0, \quad \left(\phi_k^{(f)}\right)\bigg|_{z=0} = 0, \text{ (short-circuited case)} \tag{C.4b}$$

$$\left(\phi_k^{(d)} - \phi_k^{(f)}\right)\bigg|_{z=h} = 0, \quad \left(-\varepsilon_0 \varepsilon_e \frac{\partial \phi_k^{(d)}}{\partial z} + \varepsilon_0 \varepsilon_b \frac{\partial \phi_k^{(f)}}{\partial z} - P_k\right)\bigg|_{z=h} = \sigma_k, \tag{C.4c}$$

where $\sigma_k = -\varepsilon_0 \frac{\phi_k^{(f)}(h)}{\lambda}$. Let us look for the solution of Eq.(C.3) in the form of $P_3 \sim exp(qz)$, and obtain:

$$(\tilde{\alpha} + g_{44}k^2 - g_{33}q^2)P_k = -q\phi_k^{(f)}, \tag{C.5a}$$

$$(q^2 - k^2)\phi_k^{(f)} = \frac{1}{\varepsilon_0\varepsilon_b}qP_k, \tag{C.5b}$$

where the inverse characteristic length $q$ satisfies the following equation:

$$(q^2 - k^2)(\tilde{\alpha} + g_{44}k^2 - g_{33}q^2) + \frac{q^2}{\varepsilon_0\varepsilon_b} = 0. \tag{C.6a}$$

This equation can be rewritten as:

$$q^4 - \left(\frac{\tilde{\alpha}+g_{44}k^2}{g_{33}} + k^2 + \frac{1}{\varepsilon_0\varepsilon_b g_{33}}\right)q^2 + \frac{\alpha+g_{44}k^2}{g_{33}}k^2 = 0. \tag{C.6b}$$

Since $\frac{1}{\varepsilon_0\varepsilon_b g_{33}} \gg \frac{\alpha+g_{44}k^2}{g_{33}} + k^2$ in most cases, the following approximations are valid for the solutions of Eq.(C.6):

$$q_1 \approx \pm k\sqrt{\frac{\tilde{\alpha}+g_{44}k^2}{\tilde{\alpha}+g_{44}k^2+g_{33}k^2+\frac{1}{\varepsilon_0\varepsilon_b}}} \approx \pm k\sqrt{\varepsilon_0\varepsilon_b(\tilde{\alpha} + g_{44}k^2)}, \tag{C.7a}$$

$$q_2 \approx \pm\sqrt{\frac{1}{\varepsilon_0\varepsilon_b g_{33}} - q_1^2} \approx \pm\sqrt{\frac{1}{\varepsilon_0\varepsilon_b g_{33}}}. \tag{C.7b}$$

The expressions for the amplitudes $\phi_k^{(f)}$, $\phi_k^{(d)}$ vs. the polarization $P_k$, and the wavenumber $k$ should be found from the boundary conditions (C.4). Using the functions $P_k = p_1 e^{q_1 z} + p_2 e^{-q_1 z} + p_3 e^{q_2 z} + p_4 e^{-q_2 z}$, $\phi_k^{(in)} = \phi_1 e^{q_1 z} + \phi_2 e^{-q_1 z} + \phi_3 e^{q_2 z} + \phi_4 e^{-q_2 z}$ and $\phi_k^{(out)} = g\frac{\sinh(k(d+h-z))}{\sinh(k d)}$, we derived the cumbersome characteristic equation:

$$\frac{q_1 q_2(q_1^2-q_2^2)}{(k^2-q_1^2)(k^2-q_2^2)}\left[\frac{k^2 q_2}{k^2-q_1^2}\cosh(q_1 h)\sinh(q_2 h) - \frac{k^2 q_1}{k^2-q_2^2}\cosh(q_2 h)\sinh(q_1 h) + \frac{1}{\varepsilon_b}\frac{q_1 q_2(q_1^2-q_2^2)}{(k^2-q_1^2)(k^2-q_2^2)}\left\{\varepsilon_d k \cdot \coth(d k) + \frac{1}{\lambda}\right\}\sinh(q_1 h)\sinh(q_2 h)\right] = 0. \tag{C.8a}$$

Under the assumptions $q_2 \gg |q_1|$ and $q_2 h \gg 1$, which are reasonable since the length-scale $q_2^{-1}$ is much smaller than lattice constant in most of the cases, and Pade approximation for tangent function $\tan(y) \approx y/\left(1 - \frac{4}{\pi^2}y^2\right)$ at $y^2 \lesssim \pi^2/4$, one could transform Eq.(C.8a) to the following form

$$1 + \frac{q_1^2/k^2}{1+\frac{4}{\pi^2}(q_1 h)^2}\left\{\frac{\varepsilon_d k}{\tanh(kd)} + \frac{1}{\lambda}\right\}\frac{h}{\varepsilon_b} \approx 0. \tag{C.8b}$$

Then, using the evident form of parameter $q_1$ from Eq.(C.7a), one could rewrite Eq.(C.8b) in the form

$$\tilde{\alpha} + g_{44}k^2 + \frac{1+\varepsilon_0\varepsilon_b g_{33}k^2}{\varepsilon_0\varepsilon_b}\left(1 + \left(\frac{2}{\pi}kh\right)^2 + \left\{\frac{\varepsilon_d k}{\tanh(kd)} + \frac{1}{\lambda}\right\}\frac{h}{\varepsilon_b}\right)^{-1} \approx 0. \tag{C.8c}$$

It is seen that the condition (C.8b) allows one to introduce the imaginary parameter $i\xi$

$$i\xi \equiv q_1 \approx \sqrt{-k^2 / \left\{ \left(\frac{2}{\pi}kh\right)^2 + \left(\frac{\varepsilon_d k}{\tanh(kd)} + \frac{1}{\lambda}\right)\frac{h}{\varepsilon_b} \right\}}. \qquad (C.8d)$$

It is seen that $q_1$ is purely imaginary, so that the introduced parameter $\xi$ is a positive real value, $\xi = k / \sqrt{\left(\frac{2}{\pi}kh\right)^2 + \left(\frac{\varepsilon_d k}{\tanh(kd)} + \frac{1}{\lambda}\right)\frac{h}{\varepsilon_b}}$. The parameter is a function of sizes $h$, $d$, and $\lambda$ as well as the function of the wavenumber $k$.

From the other hand, allowing for the temperature dependence of $\tilde{\alpha}$, one could consider (C.8c) as the equation for the critical temperature of the transition between the PE phase and PDFI state. This temperature depends on the wave vector $k$ and should be maximized with respect to it. In result we obtain the following expression:

$$T_{PDFI} = \tilde{T}_C - \frac{g_{44}}{\alpha_T \varepsilon_b} \frac{\pi^2}{4h} \left( \frac{4}{\pi}\sqrt{\frac{\varepsilon_b}{\varepsilon_0 g_{44}}} - \frac{\varepsilon_b}{h} - \frac{\varepsilon_d}{d} - \frac{1}{\lambda} \right), \qquad (C.8e)$$

where we introduced the Curie temperature renormalized by the mismatch strain:

$$\tilde{T}_C = T_C + \frac{2}{\alpha_T} \frac{Q_{13}(s_{22}-s_{12}) + Q_{23}(s_{11}-s_{12})}{s_{11}s_{22} - s_{12}^2} u_m. \qquad (C.8f)$$

It should be noted that the temperature of the PE-SDFI, $T_{SDFI} = \tilde{T}_C - \frac{1}{\alpha_T \varepsilon_0 \left(\varepsilon_b + \frac{h}{d}\varepsilon_d + \frac{h}{\lambda}\right)}$, is always smaller than $T_{PDFI}$ in the region $\frac{2}{\pi}\sqrt{\frac{\varepsilon_b}{\varepsilon_0 g_{44}}} > \frac{\varepsilon_b}{h} + \frac{\varepsilon_d}{d} + \frac{1}{\lambda}$ (otherwise $T_{PDFI}$ losts its meaning due to the complexity of wave vector $k_{cr}$). It is obvious that under the condition

$$\frac{2}{\pi}\sqrt{\frac{\varepsilon_b}{\varepsilon_0 g_{44}}} = \frac{\varepsilon_b}{h} + \frac{\varepsilon_d}{d} + \frac{1}{\lambda} \qquad (C.8g)$$

the transition temperatures are equal, which corresponds to the tricritical point

$$T_{trc} = T_{SDFI} = T_{PDFI} = \tilde{T}_C - \frac{g_{44}}{\alpha_T} \frac{\pi}{2h\sqrt{\varepsilon_0 \varepsilon_b g_{44}}}. \qquad (C.8h)$$

Thus the conditions (C.8g) and (C.8h) define the tricritical point, where all three phases are in equilibrium.

### C2. Direct variational method for the domain structure consideration

Considering homogeneous and harmonic periodic solutions, one can use the trial functions in the following form:

$$P_3(x,z) = P(z) + 2A \cos(kx) \left\{ \cos(\xi z) + \frac{\xi \sin(\xi h)}{\mu \sinh(\mu h)} \cosh(\mu z) \right\}, \qquad (C.9a)$$

$$\varphi_f(x,z) = \frac{1}{\varepsilon_0 \varepsilon_b} \int_0^z P(\tilde{z}) d\tilde{z} - z \left\{ \frac{\varepsilon_d \left(\frac{d}{\lambda \varepsilon_d} + 1\right)}{d\left(\frac{1}{\lambda} + \frac{\varepsilon_b}{h} + \frac{\varepsilon_d}{d}\right)} \frac{1}{\varepsilon_0 \varepsilon_b h} \int_0^h P(\tilde{z}) d\tilde{z} \right\} + \sum_k \phi_k^{(f)}(z) \exp(ikx). \qquad (C.9b)$$

Here we introduced parameter $\mu = 1/\sqrt{\varepsilon_0 \varepsilon_b g_{33}}$. The first terms are a homogeneous solution with the magnitude "$P$", corresponding to the single domain state. The last terms are a harmonically modulated solution with amplitude "$A$" proportional to the maximal polarization of the polydomain state.

Constant $\xi$ are given by Eq. (C.8d). In the case of constant $P$, the functions (C.9) satisfy the boundary conditions (C.4a) and (C.4b) for arbitrary values of wave number $k$, and the conditions (C.4c) are satisfied under the condition (C.8). In this case Eqs.(C.9) can be rewritten as follows:

$$P_3 \approx P + 2A\cos(kx)\left[\cos(\xi z) + \frac{\xi \sin(\xi h)}{\mu \sinh(\mu h)}\cosh(\mu z)\right], \quad \text{(C.10a)}$$

$$\varphi_f \approx \frac{zP}{\varepsilon_0}\frac{1}{h\left(\frac{1}{\lambda}+\frac{\varepsilon_b}{h}+\frac{\varepsilon_d}{d}\right)} + \frac{2A}{\varepsilon_0\varepsilon_b}\cos(kx)\left[\frac{\xi\sin(\xi z)}{\xi^2+k^2} + \frac{\xi\sin(\xi h)}{\sinh(\mu h)}\frac{\sinh(\mu z)}{\mu^2-k^2}\right]. \quad \text{(C.10b)}$$

The amplitudes "$P$" and "$A$" are the variational parameters which could be determined from the minimization of the free energy functional. In the most cases the terms proportional to $\exp(-\mu z)$ give only the slightest corrections to the average values and can be omitted, while their presence is important for the fulfillment of the boundary conditions. Keeping these arguments in mind, one could substitute Eqs. (C.10) to the free energy functional:

$$G = \int_{-\infty}^{\infty} dx \int_0^h dz \left\{\frac{\tilde{\alpha}}{2}P_3^2 + \frac{\tilde{\beta}}{4}P_3^4 + \frac{\tilde{\gamma}}{6}P_3^6 + \frac{\tilde{\delta}}{8}P_3^8 - E_{ext}P_3 + \frac{g_{33}}{2}\left(\frac{\partial P_3}{\partial z}\right)^2 + \frac{g_{44}}{2}\left(\frac{\partial P_3}{\partial x}\right)^2 + \frac{1}{2}P_3\frac{\partial \phi}{\partial z}\right\}.$$

(C.11a)

and obtain the following expression:

$$G[P,A] \approx \frac{\alpha_p}{2}P^2 + \frac{\alpha_a}{2}A^2 + \frac{\tilde{\beta}}{4}\left(P^4 + 6P^2A^2 + \frac{9}{4}A^4\right) + \frac{\tilde{\gamma}}{6}(P^6 + 15P^4A^2 + \mathcal{O}[A^4]P^2 + \mathcal{O}[A^6]) +$$

$$\frac{\tilde{\delta}}{8}(P^8 + 28P^6A^2 + \mathcal{O}[A^4]P^4 + \mathcal{O}[A^6]P^2 + \mathcal{O}[A^8]) - E_{eff}P. \quad \text{(C.11b)}$$

Here we introduced the renormalized coefficients and effective electric field:

$$\alpha_p = \tilde{\alpha} + \frac{1}{\varepsilon_0 h\left(\frac{1}{\lambda}+\frac{\varepsilon_b}{h}+\frac{\varepsilon_d}{d}\right)}, \quad \text{(C.12a)}$$

$$\alpha_a = \tilde{\alpha} + g_{33}\xi^2 + g_{44}k^2 + \frac{1}{\varepsilon_0\varepsilon_b}\frac{\xi^2}{k^2+\xi^2} \approx \tilde{\alpha} + g_{44}k^2 + \frac{1}{\varepsilon_0\varepsilon_b}\frac{1+\varepsilon_0\varepsilon_b g_{33}k^2}{1+\left(\frac{2}{\pi}kh\right)^2 + \left\{\frac{\varepsilon_d k}{\tanh(kd)}+\frac{1}{\lambda}\right\}\frac{h}{\varepsilon_b}}, \quad \text{(C.12b)}$$

$$E_{eff} = -\frac{\varepsilon_d U}{d\, h\left(\frac{1}{\lambda}+\frac{\varepsilon_b}{h}+\frac{\varepsilon_d}{d}\right)}. \quad \text{(C.12c)}$$

In Eq.(C.12b) we used approximation for $\xi^2$ from Eq.(C.8d). Equations of state can be obtained from the minimization of (C.11b) over $P$ and $A$. For small $A$ all terms proportional to $\mathcal{O}[A^4]P^2$ and higher can be neglected and the equations of state acquires the form:

$$\left(\alpha_p + 3\tilde{\beta}A^2\right)P + \left(\tilde{\beta} + 10\tilde{\gamma}A^2\right)P^3 + \left(\tilde{\gamma} + 42\tilde{\delta}A^2\right)P^5 + \tilde{\delta}P^7 = E_{eff}, \quad \text{(C.13a)}$$

$$\left(\alpha_a + 3\tilde{\beta}\,P^2\right)A + \frac{9}{4}\tilde{\beta}\,A^3 = 0. \quad \text{(C.13b)}$$

The solution of the system (C.13) contains the following phases, which are described below:

I. The paraelectric (PE) phase with $A = 0$ and $P \cong \frac{E_{eff}}{\alpha_p}$. Here $P = 0$ at $E_{eff} = 0$.

II. The single domain ferrielectric (SDFI) phase with $A = 0$ and $P \neq 0$ at $E_{eff} = 0$.

III. The "pure" polydomain ferroelectric (PDFE) phase $A \neq 0$ and $P = 0$ at $E_{eff} = 0$.

IV. The "rippled" polydomain modulated ferrielectric (MFE) phase, with $A \neq 0$ and $P \neq 0$ at $E_{eff} = 0$.

The concrete phase is stable when the matrix, listed below, is positively defined in the phase

$$\widehat{M} = \begin{pmatrix} \frac{\partial^2 G}{\partial P^2} & \frac{\partial^2 G}{\partial P \partial A} \\ \frac{\partial^2 G}{\partial P \partial A} & \frac{\partial^2 G}{\partial A^2} \end{pmatrix}. \tag{C.14}$$

The matrix of generalized dielectric susceptibilities is equal to the inverse matrix of the second derivatives:

$$\hat{\chi} = \begin{pmatrix} \frac{\partial^2 G}{\partial P^2} & \frac{\partial^2 G}{\partial P \partial A} \\ \frac{\partial^2 G}{\partial P \partial A} & \frac{\partial^2 G}{\partial A^2} \end{pmatrix}^{-1}. \tag{C.15}$$

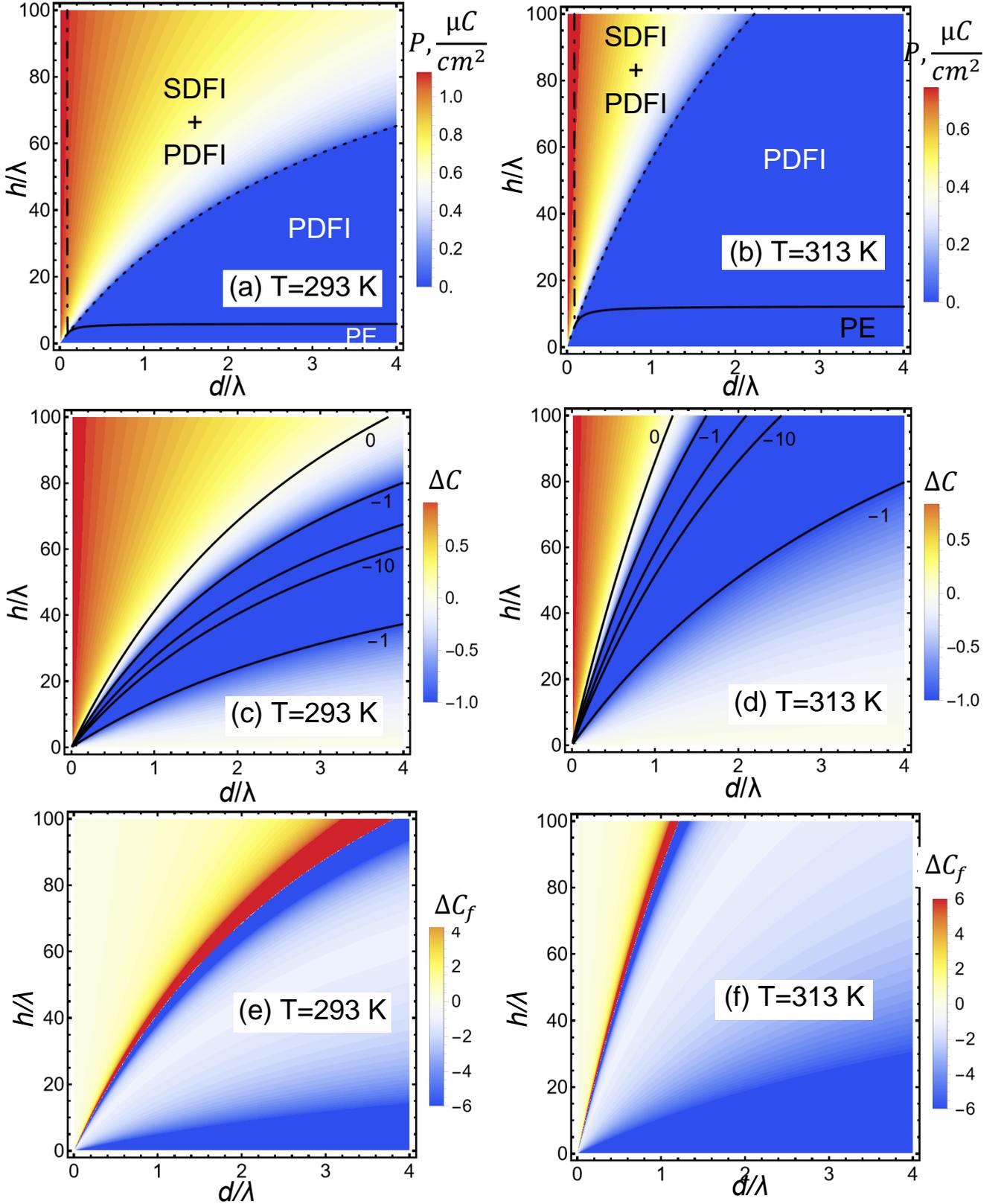

**Figure D1.** Color maps showing the dependence of the spontaneous polarization (**a, b**), relative capacitance $\Delta C$ (**c, d**), and the apparent capacitance of the CIPS layer $\Delta C_f$ (**e, f**) on the relative thickness $h/\lambda$ of the CIPS film and relative thickness $d/\lambda$ of the SiO$_2$ layer for the temperatures, $T$=293 K (**a, c, e**) and 323 K (**b, d, f**).. Misfit strain is $u_m = 0.3\%$, $\varepsilon_d = 3.9$.

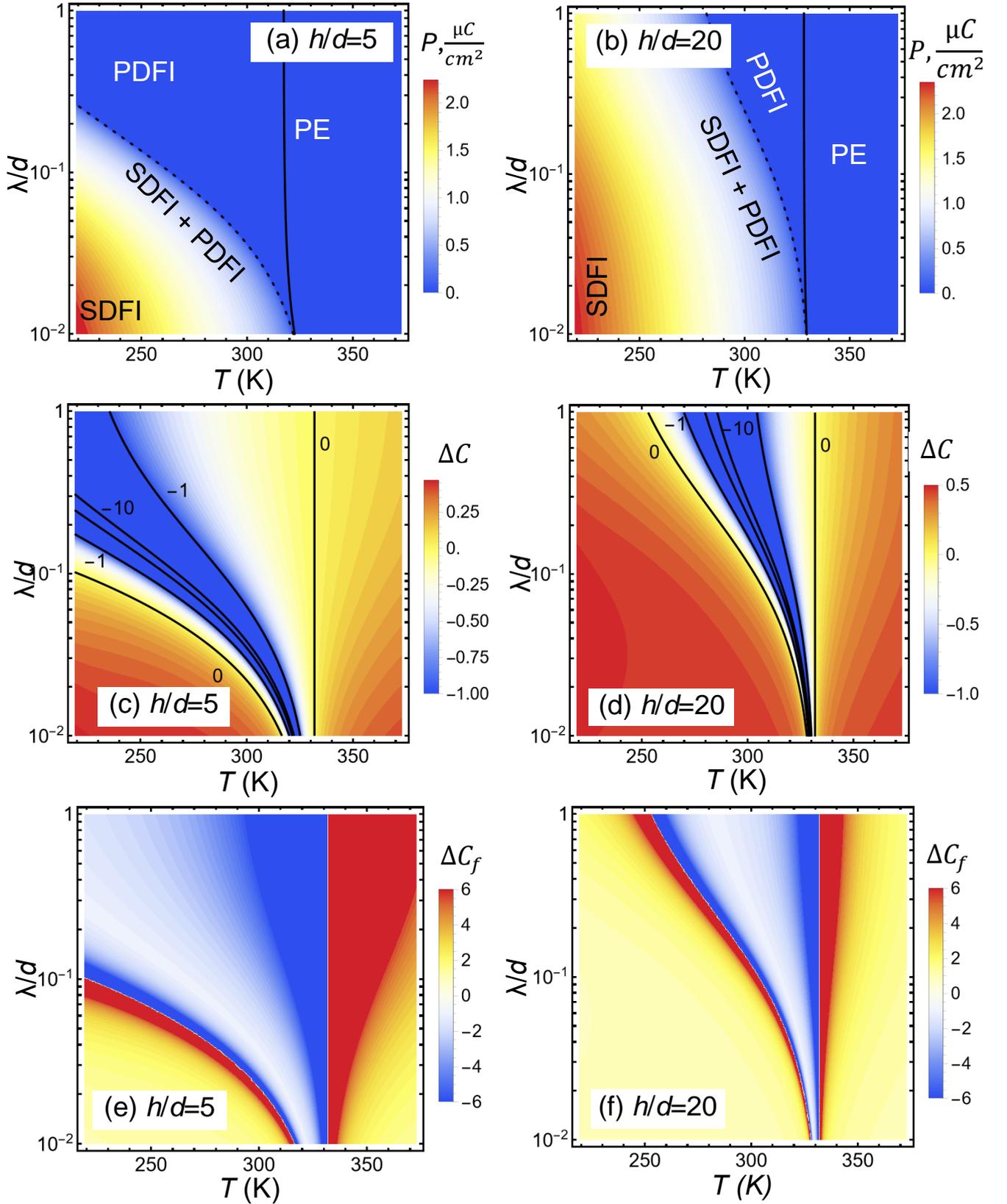

**Figure D2.** Color maps showing the dependence of the spontaneous polarization (**a, b**), relative capacitance $\Delta C$ (**c, d**), and the apparent capacitance of the CIPS layer $\Delta C_f$ (**e, f**) on the dimensionless screening length $\frac{\lambda}{d}$ of the 2D-MoS$_2$ layer and temperature $T$ for different values of the ratio $\frac{h}{d}=5$ (**a, c, e**) and $\frac{h}{d}=20$ (**b, d, f**). Misfit strain is $u_m = 0.3\%$, $\varepsilon_d = 3.9$.

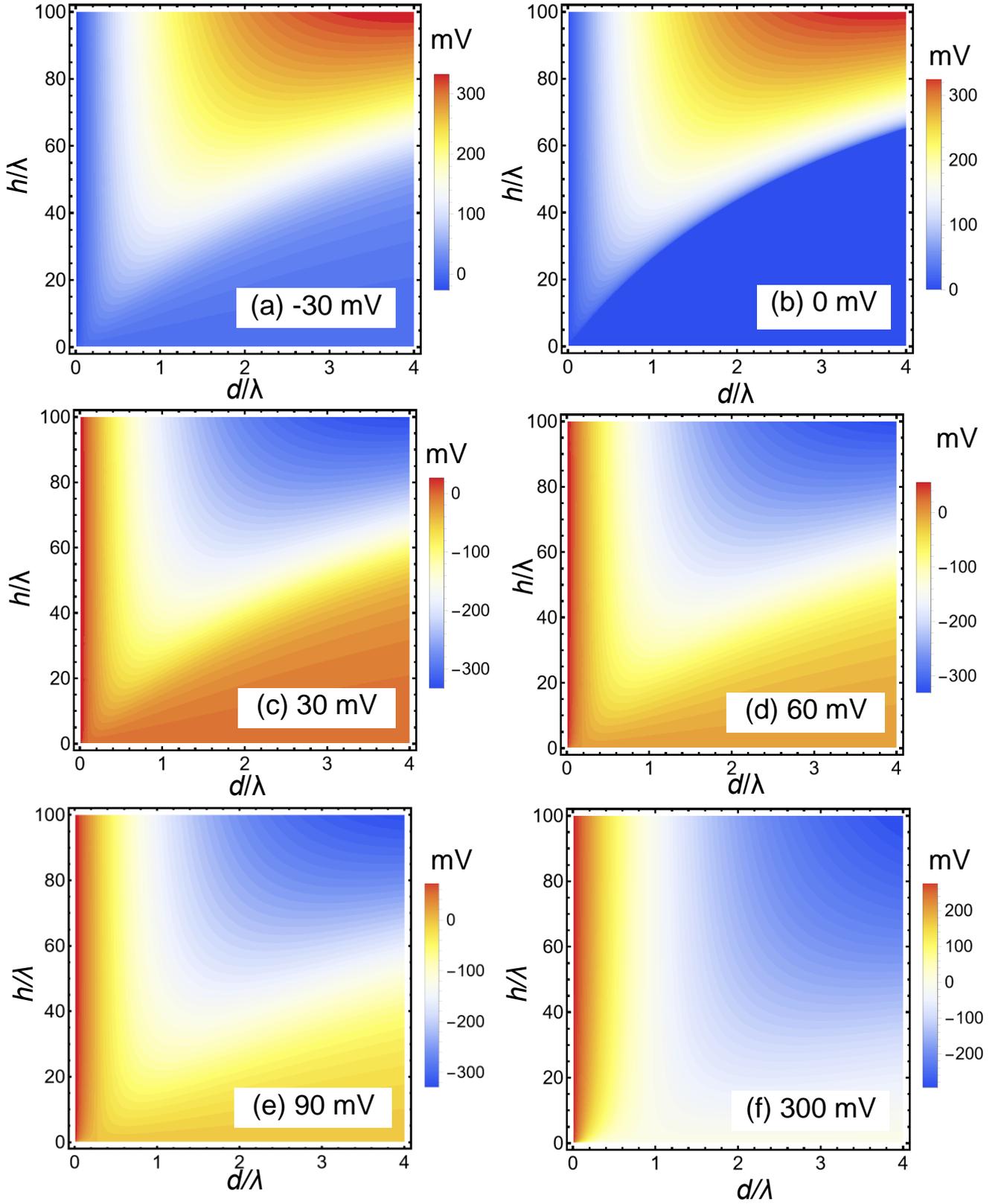

**Figure D3.** The dependence of the electrostatic potential at the interface, $\varphi_f(h)$, on the relative thickness $h/\lambda$ of the CIPS film and relative thickness $d/\lambda$ of the SiO$_2$ layer calculated at the temperature $T$=293 K and different values of applied voltage, $U$ =-30 mV (**a**), 0 V (**b**), 30 mV (**c**), 60 mV (**d**), 90 mV (**e**) and 300 mV (**f**). Misfit strain is $u_m = 0.3\%$, $\varepsilon_d = 3.9$.

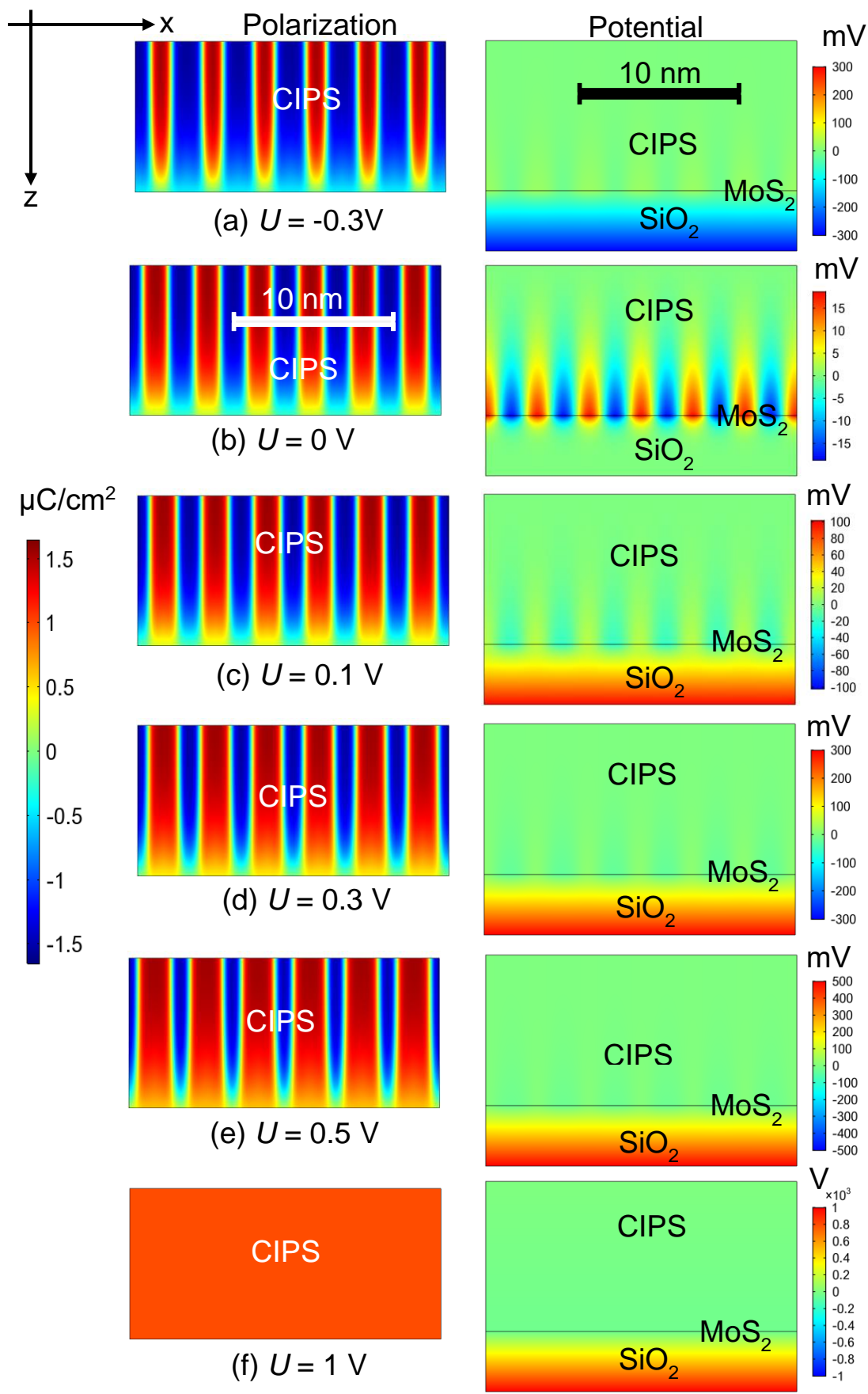

**Figure D4**. Distributions of the CIPS polarization (left column) and the electrostatic potential inside the structure "CIPS – 2D-MoS₂ – SiO₂" (right column) calculated for different values of applied voltage $U = $ -0.3

V **(a)**, 0 V **(b)**, 0.1 V **(c)**, 0.3 V **(d)**, 0.5 V **(e)** and 1 V **(f)**. Parameters: $h = 10$ nm, $\lambda$=0.5 nm, $d = 4$ nm, $\varepsilon_d = 3.9$, $u_m = 0.3\%$, $T = 293$ K.